\documentclass[twocolumn,groupedaddress,amsmath,amssymb,prb,aps,longbibliography]{revtex4-1}

\usepackage{float}
\usepackage{siunitx}
\usepackage{graphicx}
\usepackage{dcolumn}
\usepackage{bm}
\usepackage{booktabs}
\usepackage{multirow}

\usepackage{rotating}
\usepackage{xcolor}

\begin{document}

\title{Effective dynamic constants for nonequilibrium third-principles
  simulations}

\author{Mauro Pulzone,$^{1,2}$ I\~nigo Robredo-Magro,$^{1}$ and Jorge
  \'I\~niguez-Gonz\'alez$^{1,2}$}

\affiliation{
  \mbox{$^{1}$Luxembourg Institute of Science and Technology (LIST),
    Avenue des Hauts-Fourneaux 5, L-4362 Esch/Alzette, Luxembourg}\\
 \mbox{$^{2}$Department of Physics and Materials Science, University
   of Luxembourg, Rue du Brill 41, L-4422 Belvaux, Luxembourg}}

\begin{abstract}
Computational studies of the thermodynamic properties of materials at
the mesoscopic and macroscopic scales -- involving lengths and times
of at least $\mu$m and $\mu$s, respectively -- rely on a
coarse-graining approximation such that only a few relevant collective
variables are treated explicitly. Those variables typically take the
form of fields defined everywhere in space (e.g., a polarization field
in dielectric or ferroelectric media) or macroscopic quantities when
spatial inhomogeneities can be treated implicitly (e.g., the volume
average of the electric polarization). The free energy is usually
expressed as a Landau-like potential whose temperature-dependent
minima track stable states, characteristic equilibrium fluctuations
being implicitly accounted for. Further, the response of the system to
external perturbations, and its relaxation toward thermal equilibrium,
are described in terms of simple equations of motion governed by
effective inertial and viscous-damping constants. There is
considerable literature on the problem of deriving Landau free energy
potentials, from either experiment or predictive atomistic
simulations, including recent efforts to develop systematic
machine-learning approaches that we denote ``third principles''. Much
less attention has received the calculation of the effective constants
controlling the nonequilibrium macroscopic or mesoscopic
dynamics. Here we tackle that problem, describing a protocol that
allows us to compute the temperature-dependent inertial and damping
coefficients associated to the electric polarization in representative
soft-mode ferroelectric PbTiO$_{3}$. Our scheme lends itself to a
widespread application, although the non-trivial behaviors found in
PbTiO$_{3}$ suggest that more case studies will be needed to finetune
a general and robust calculation protocol. Our results also allow us
to comment on common assumptions in the literature of effective
dynamic treatments of ferroelectrics and related materials.
\end{abstract}

\maketitle


\section{Introduction}

The advent of machine-learned interatomic potentials (MLIPs) has
ushered in a golden age of atomistic simulation, already covering, in
essence, all structural and lattice-dynamical properties of arbitrary
materials in the nanoscale ($\sim$nm,
$\sim$ns).\cite{bartok15,wang18,jinnouchi19,fan21,musaelian23,xie23,batatia24,jacobs25,cheng25,fan26}
With increasing evidence that MLIPs can be used to unveil non-trivial
behaviors out of the scope of the training set\cite{robredomagro26a}
and exciting prospects for incorporating magnetic and electronic
phenomena,\cite{zhong23} expectations from predictive atomistic
simulations are constantly being redefined, expanding quickly.

Yet, MLIPs are far from providing a solution to all problems of
interest. For example, whenever we deal with mesoscale or macroscopic
phenomena, we are typically interested in length and time scales of at
least $\mu$m and $\mu$s, respectively, where simulations based on
atomistic methods are unfeasible because of their high computational
cost. Theoretical approaches to such problems typically rely on
continuum field theories, and even on macroscopic models whenever
spatial inhomogeneities can be ignored. Of particular interest are
cases involving ferroic materials (ferromagnetic, ferroelectric,
ferroelastic, and related systems) that often present complex
mesoscale states with non-trivial topologies. Remarkably, many such
cases can be treated by relatively simple Landau-like theories,
yielding qualitative and quasi-quantitatively satisfactory
results.\cite{hu98,chen08,toledano-book1987,chandra07,wen22}

Studies based on Landau-like theories comprise two basic
elements. First, there is a free energy potential expressed as a
function (functional) of the relevant variables (fields). To fix
ideas, consider the case of ferroelectric materials, which exhibit
strongly anharmonic thermodynamic behavior, from phase transitions and
complex textures to anomalous and highly tunable responses, and offer
a stringent test.\cite{lines-book1977,rabe-book2007} In this context,
we work with a Landau (Ginzburg–Landau) potential that depends on the
macroscopic electric polarization (polarization field) and whose
minima define the possible stable equilibrium states and their
temperature evolution. (In what follows, we use the term ``Landau
potential'' to refer to both Landau and Ginzburg–Landau models,
distinguishing between them only when needed for clarity or emphasis.)
Such potentials have enabled numerous successful studies predicting or
helping explain novel behaviors in nanostructured ferroelectrics, from
chiral topological states to negative-capacitance
responses.\cite{yadav16,damodaran17,hong17,das19,zubko16,yadav19,das21,junquera23,lukyanchuk25}
Many of these emerging phenomena stem from complex ferroelectric
multidomain states, with domain walls often playing a key
role.\cite{wojdel14a,zubko16,goncalves19} While atomistic simulations
have been essential for uncovering certain key features and helping
derive Ginzburg–Landau potentials that retain sufficient
detail,\cite{wojdel14a,zubko16} the most interesting phenomena are
clearly mesoscopic in nature, making field theory the approach of
choice. A considerable number of studies have addressed the problem of
deriving free energy potentials from
experiment\cite{devonshire54,haun87} or more fundamental atomistic
simulations.\cite{iniguez01,kumar10,pulzone25,durdiev25} Recent
efforts, including systematic perturbative and machine-learning
approaches,\cite{pulzone25,schiaffino17,shapovalov23} show promise for
capturing increasingly intricate effects in a predictive {\it ab
  initio} manner. Hence, we are confident that the reliable
calculation of Landau and Ginzburg-Landau potentials for general and
potentially complex materials, based solely on first-principles data,
is well underway.

Second, to go beyond quasi-static properties, we must extend the
theory by introducing suitable equations of motion. It is worth
stressing that the natural thermal oscillations of the system around
equilibrium are already captured, implicitly, in the Landau potential
itself; they can therefore be regarded as quasi-static and, to some
extent, trivial. The real challenge lies in describing nonequilibrium
dynamics at the mesoscopic or macroscopic scales. For example, imagine
subjecting the material to an external stimulus that takes it away
from its initial equilibrium state. We may then be interested in
characterizing how it evolves toward a new equilibrium configuration
(as it would, e.g., when a transition is triggered by a strong
electric field pulse), how it reaches a stationary state (e.g., when
driven by an alternating field constantly applied), or how it relaxes
back to the initial state (e.g., following a relatively weak field
pulse). Dynamic phenomena of this kind are attracting increasing
attention, in particular in the ferroelectrics community: for
instance, in studies of photovoltaic effects, ultrafast control with
lasers, engineering dynamical nonlinear responses for neuromorphic
applications, or dynamically-induced magnetic
fields.\cite{yang20,prosandeev21,li21b,chen22,wang24,wang25} However,
most existing time-dependent Landau-like treatments rely on
phenomenological dynamic constants and largely untested
approximations. Despite some remarkable
efforts,\cite{thomas10,hlinka13,liu21,durdiev25} we still lack a
systematic {\it ab initio} approach to this problem.

To better frame the question, let us recall the equations of motion
associated to the Landau potential $F(\bm{P})$ of a ferroelectric
compound. They are typically written as
\begin{equation}
      \mu_{\alpha} \ddot{P}_{\alpha} + c_{\alpha}
      \dot{P}_{\alpha} = -\frac{\partial F}{\partial P_{\alpha}}
      \label{eq:lk}
\end{equation}
for each polarization component $P_{\alpha}$. Here, $\mu_{\alpha}$ and
$c_{\alpha}$ are inertial and viscous-damping constants, respectively;
dots stand for time derivatives and the term on the right-hand side
can be seen as a generalized force acting on $P_{\alpha}$. The
corresponding time-dependent Ginzburg-Landau equations are obtained by
writing the polarization field $\bm{P}(\bm{r})$ in place of $\bm{P}$
and computing the associated position-dependent force field as a
functional derivative of $F[\bm{P}(\bm{r})]$.

In the vast majority of studies, the time propagation in
Eq.~(\ref{eq:lk}) is computed in the relaxational regime, thus making
$\mu_{\alpha} = 0$. The simplified equations are often called
``Landau-Khalatnikov''. Further, $c_{\alpha} = 1$ is commonly assumed,
the simulation thus reducing, in essence, to a steepest-descent
relaxation toward a local minimum of the free energy. However,
triggered by the mentioned interest in dynamical and ultrafast
phenomena, the need to account for inertial effects is now
increasingly recognized, and so is the need to use physically
meaningful values of the dynamic constants $\mu_{\alpha}$ and
$c_{\alpha}$.

In the context of ferroelectrics, reasonable estimates for the
inertial and damping parameters controlling the polarization dynamics
can often be derived from experimental information, e.g. by fitting to
spectroscopic data or by trying to reproduce observed
polarization-switching
behavior.\cite{luspin80,fontana91,foster93a,foster93b,cho01,hlinka13}
However, such empirical constants are not available in the general
case. Difficulties arise when we need to consider order parameters
that are not easily probed experimentally (e.g., structural modes away
from the $\Gamma$-point of the Brillouin zone) or when dealing with
virtual materials not yet fabricated. In this context, we must
highlight works like those of Liu {\it et al.}\cite{liu21} or Durdiev
{\it et al.},\cite{durdiev25} which constitute, to the best of our
knowledge, the only existing attempts at calculating the constants
controlling ferroelectric polarization dynamics from atomistic
molecular-dynamics (MD) simulations.

Here we present our own take on this problem. We introduce an approach
to compute the inertial and damping constants effectively controlling
the dynamics of the macroscopic electric polarization, and we
illustrate the method with an application to representative compound
PbTiO$_{3}$. We show that it is possible to rely exclusively on the
atomistic dynamics at equilibrium, a convenient simplification
inspired by the usual approach to derive phonon frequencies and
associated damping constants from experimental spectroscopic data. We
discuss the difficulties that arise owing to the anharmonic couplings
that are ubiquitous in soft-mode ferroelectrics and related compounds,
and we introduce a practical strategy to tackle them. Finally, we
discuss the implications of our results, placing them in the broader
context of the ongoing effort toward quantitative and predictive
``third-principles'' simulations of dynamic phenomena at the
mesoscopic and macroscopic scales.

\section{Our approach}\label{sec:approach}

We now introduce a basic protocol for calculating effective dynamic
constants for the polarization by fitting the dielectric power
spectrum. We first present the methodology in general terms, noting
that it parallels standard experimental procedures used in vibrational
spectroscopy
studies.\cite{luspin80,fontana91,foster93a,foster93b,cho01,hlinka13}
Then, we justify our choice of PbTiO$_{3}$ as a suitably challenging
example of application and discuss briefly the expectations one may
have about its power spectrum {\it a priori}. Our discussion applies
to any material displaying soft phonon modes, and thus posing a
non-trivial test in the present context. We also introduce the
atomistic ``second-principles'' model used in our MD
simulations,\cite{wojdel13} giving the necessary technical details.

For pedagogical purposes, some of the arguments below are adapted to
the illustrative example of PbTiO$_{3}$, where the goal is to derive
effective equations of motion for the three spatial components of the
order parameter $\bm{P}(t)$, which accounts for the material's
macroscopic polarization. Application to other cases should be
relatively straightforward and is addressed in
Section~\ref{sec:discussion}.

\subsection{Modeling the equilibrium power spectrum}
\label{sec:method}

Imagine we can compute (or measure) the time evolution $\bm{P}_{\rm
  tot}(t)$ of the polarization of a material at equilibrium. Note that
here we distinguish between the total polarization of the material
$\bm{P}_{\rm tot}(t)$, as provided by experiment or atomistic
simulation, and the polarization $\bm{P}(t)$ that appears in a
time-dependent Landau-like treatment. We will be more specific about
this difference below.

To study the equilibrium oscillations of $\bm{P}_{\rm tot}(t)$, it is
convenient to work with $\Delta \bm{P}_{\rm tot}(t) = \bm{P}_{\rm
  tot}(t) - \bm{P}_{\rm tot, eq}$, where the equilibrium polarization
$\bm{P}_{\rm tot, eq} = \langle \bm{P}_{\rm tot}(t) \rangle$ is a time
average. Critically, we will assume throughout that $\Delta\bm{P}_{\rm
  tot}(t)$ is relatively small in magnitude, so a harmonic treatment
of the associated energy variations around $\bm{P}_{\rm tot, eq}$ is
warranted.

We define the polarization spectrum $\Delta\bm{P}_{\rm tot}(\omega)$ as
\begin{equation}
  \Delta\bm{P}_{\rm tot}(\omega) = \int_{0}^{\infty} \Delta\bm{P}_{\rm
    tot}(t) \exp{(i\omega t)} dt \; .
\end{equation}
We also introduce the power spectrum of the polarization, defined as
$S_{{\rm tot},\alpha}(\omega) = |\Delta P_{{\rm tot},
  \alpha}(\omega)|^{2}$ for each spatial component $\alpha$. We now
discuss how to model $\bm{S}_{\rm tot}(\omega)$ to extract the dynamic
constants pertaining to the polarization $\bm{P}(t)$ that we want to
describe in our Landau-like approach.

\subsubsection{Equations of motion}

Ideally, we would like to use a simple equation of motion, as that in
Eq.~(\ref{eq:lk}), to describe the time-dependent function
$\Delta\bm{P}_{\rm tot}(t)$. However, it is important to recognize
that the total polarization is, in all but the simplest of cases, a
collective variable that can be split into different contributions,
each having distinct spectral signatures. Note that all polar phonons
contribute to $\Delta\bm{P}_{\rm tot}(t)$. Further, various anharmonic
and potentially complex effects -- from the so-called overtones to
order-disorder features in the dynamics of certain oscillators -- may
contribute to $\Delta\bm{P}_{\rm tot}(t)$ as well. Hence, let us write
\begin{equation}
  \Delta\bm{P}_{\rm tot}(t) = \sum_{i} \Delta \bm{P}_{i}(t) \; ,
  \label{eq:multimode}
\end{equation}
where each term $\Delta \bm{P}_{i}(t)$ quantifies a distinct
contribution to the dynamics of the total polarization. For the sake
of a clear presentation, let us assume for the time being that these
$\Delta \bm{P}_{i}$'s are associated to polar phonons. Let us also
introduce the notation
\begin{equation}
  \Delta P_{i\alpha}(t) = x_{i\alpha} \Delta P_{i}(t) \; ,
\end{equation}
to account for the fact that only the amplitude $\Delta P_{i}(t)$ is
time dependent, while the relative contributions $x_{i\alpha}$ to the
polarization along the different directions can be taken as
constant. (The $x_{i\alpha}$ factors may depend on $T$, though.) In
analogy with the textbook perturbative treatment for normal modes, we
introduce the following equation of motion for each individual $\Delta
P_{i}(t)$
\begin{equation}
  \Delta \ddot{P}_{i} + \gamma_{i} \Delta \dot{P}_{i} + \omega_{i}^{2}
  \Delta P_{i} = 0 \; ,
  \label{eq:EoM-i}
\end{equation}
where $\omega_{i}$ is the characteristic frequency of mode $i$ and
$\gamma_{i}$ a damping constant that quantifies how quickly this mode
equilibrates. The damping constant thus captures, effectively, the
anharmonic couplings of oscillator $i$ with all the degrees of freedom
in the material, which form what is usually called a {\it thermal
  bath}. The critical assumption here is that we can decouple the
equations of motion of the individual $\Delta \bm{P}_{i}(t)$
contributions to $\Delta\bm{P}_{\rm tot}(t)$, which essentially
amounts to the following approximation for the total power spectrum:
\begin{equation}
  \bm{S}_{\rm tot}(\omega) \approx \sum_{i} \bm{S}_{i}(\omega)\; ,
  \label{eq:S}
\end{equation}
where $\bm{S}_{i}(\omega) = |\Delta\bm{P}_{i}(\omega)|^{2}$ and
\begin{equation}
  \Delta\bm{P}_{i}(\omega) = \int_{0}^{\infty} \Delta\bm{P}_{i}(t)
  \exp{(i\omega t)} dt \; .
\end{equation}
We can trivially write these quantities in terms of the fundamental
time-dependent amplitudes in Eq.~(\ref{eq:EoM-i}), by introducing
$\Delta P_{i\alpha}(\omega) = x_{i\alpha} \Delta P_{i}(\omega)$ and
$S_{i\alpha}(\omega) = x_{i\alpha}^{2}S_{i}(\omega)$.

These decoupled equations of motion are well justified if we work with
a collection of ideal harmonic oscillators. Here we assume that we can
use them in more complex scenarios as well. In particular, we suppose
that we can assign a $\Delta\bm{P}_{i}(t)$ term, with dynamics given
by Eq.~(\ref{eq:EoM-i}), to any distinct feature in the power
spectrum, be it a simple underdamped normal mode, an intricate
multi-phonon overtone, or anything else. Within this approach, all
non-trivial couplings are effectively captured in the fitting
parameters $\omega_{i}$ and $\gamma_{i}$. While such an implicit
treatment of dynamic correlations will not allow for an explicit
discussion of all underlying physical mechanisms, we deem it justified
in the present context where the aim is to obtain simple effective
theories.

\subsubsection{Contributions from underdamped oscillators}

We now discuss the form of $\bm{S}_{i}(\omega)$. Let us first consider
the case of well-behaved polar phonons, i.e., underdamped oscillators
where $\Delta \bm{P}_{i}(t)$ displays an oscillatory time dependence
resulting in a peak-like spectral feature. For such modes, one can
propose the solution
\begin{equation}
  \Delta P_{i}(t) = A_{i} \exp{\left( -\frac{\gamma_{i}}{2}t
    \right)}\, \cos{(\bar{\omega}_{i}t+\phi_{i})} \; ,
  \label{eq:Pt-underdamped}
\end{equation}
where
\begin{equation}
\bar{\omega}_{i} =
\sqrt{\omega_{i}^{2}-\left(\frac{\gamma_{i}}{2}\right)^{2}}
\label{eq:baromega}
\end{equation}
is the so-called {\it damped frequency}. (To have an underdamped
solution we need $\gamma_{i}<2\omega_{i}$.) $A_{i}$ is the amplitude
of the oscillation (which reflects phonon population) and $\phi_{i}$ a
phase factor (essentially arbitrary, fixed by the conditions at
$t=0$). The spectral function $\Delta P_{i}(\omega)$ is obtained by
Fourier transforming this solution:
\begin{equation}
\begin{split}
  \Delta P_{i}(\omega) & = A_{i} \times \\ & \left(
  \frac{\exp{(-i\phi_{i})}}{\gamma_{i} + 2i(\omega +
    \bar{\omega}_{i})} + \frac{\exp{(i\phi_{i})}}{\gamma_{i} +
    2i(\omega - \bar{\omega}_{i})} \right) \; .
\end{split}
\end{equation}
The corresponding contribution to the power spectrum, $S_{i}(\omega)$,
is an intricate function. However, if we assume that the resonant term
-- i.e., that with $(\omega - \bar{\omega}_{i})$ in the denominator --
dominates, we obtain the relatively simple expression
\begin{equation}
 S_{i}(\omega) = |\Delta P_{i}(\omega)|^{2} \approx
 \frac{A^{2}_{i}}{\gamma_{i}^{2}+ 4 (\omega - \bar{\omega}_{i})^{2}}
 \; ,
 \label{eq:Si}
\end{equation}
which does not depend on the phase $\phi_{i}$. This is the basic
Lorentzian function, of width $\gamma_{i}$ and centered around
$\bar{\omega}_{i}$, used in this work to model spectral peaks
corresponding to underdamped oscillators. We tried refinements of this
function, but they yielded essentially identical results, not
justifying the additional complexity.

\subsubsection{Overdamped modes and other features}

The dielectric spectrum will generally display features that are not
associated to a simple underdamped polar phonon. Ferroelectrics like
PbTiO$_{3}$ are very challenging from this point of view, as they
present soft phonons whose frequency is strongly temperature dependent
and, critically, becomes very small in the vicinity of the Curie
point. If we have a polar mode whose bare frequency
$\omega_{i}\rightarrow 0$, it is all but guaranteed we will at some
point enter the overdamped regime ($\gamma_{i} > 2\omega_{i}$) where
the damped frequency $\bar{\omega}_{i}$ becomes imaginary (see
Eq.~(\ref{eq:baromega})). The contribution of an overdamped phonon to
the power spectrum is purely relaxational and takes the form of what
is usually called a {\it central peak} (cp). To model such features we
assume a time dependence of the form
\begin{equation}
  \Delta P_{\rm cp}(t) = A_{\rm cp} \exp{\left( -\gamma_{\rm cp}t
    \right)}\; ,
\end{equation}
which solves Eq.~(\ref{eq:EoM-i}) for $\omega_{i} = \omega_{\rm cp}
\rightarrow 0$. The corresponding contribution to the power spectrum
is
\begin{equation}
S_{\rm cp}(\omega) = \frac{A^{2}_{\rm cp}}{\gamma_{\rm cp}^{2} +
  \omega^{2} } \; .
\label{eq:cp}
\end{equation}
As we will see, in our simulations for PbTiO$_{3}$ we find softening
phonons whose damped frequency $\bar{\omega}_{i}$ becomes very
small. We also find cp-like features probably related to such
overdamped modes, potentially compounded with other non-trivial
dynamical characteristics (e.g., of the order-disorder type) that the
system may present. We circunvent such subtleties, which are
essentially impossible to elucicate from the $\bm{P}_{\rm tot}(t)$
trajectory alone, by including $S_{\rm cp}$-like contributions to fit
the lowest-frequency part of $S_{{\rm tot},\alpha}(\omega)$ whenever
necessary.

An additional complication found in our numerical application pertains
to the presence of multi-phonon overtones that may have a rather
involved anharmonic origin, as we will see. We find it possible (and
convenient) to treat the overtones as additional underdamped
oscillators that provide their own contribution (given by an
expression as in Eq.~(\ref{eq:Si})) to the total power spectrum.

\subsubsection{Zooming in on the low-frequency polarization}

We have just described our approach to model the full dielectric power
spectrum $\bm{S}_{\rm tot}(\omega)$, which involves the fitting of a
suitable model function (Eq.~(\ref{eq:S})) that we construct by adding
contributions from underdamped polar phonons, overtones, overdamped
modes, etc. However, when developing an effective dynamic theory, we
will be typically interested in the low-frequency oscillators
alone. In the specific case of a soft-mode ferroelectric, such modes
will be associated to the development of the spontaneous polarization,
and they will also dominate the low-frequency (and static) dielectric
response. In the following we discuss how to extract such relevant
information from our analysis.

Indeed, here we want to develop ``third-principles'' models where the
dynamics of the order parameter is given by equations of motion as
those in Eq.~(\ref{eq:lk}). To do this, we first need to identify
three specific modes $P_{i}(t)$ corresponding to the three
polarization components $P_{\alpha}(t)$ that will appear in such a
theory. In the general case, we proceed as follows. For each direction
$\alpha$, we inspect the computed $S_{{\rm tot},\alpha}$ and identify
the lowest-frequency salient feature in the spectrum. Let us use the
notation $i = s\alpha$ to denote the oscillator thus selected for
direction $\alpha$, where $s$ stands for {\it soft}. In principle, the
corresponding functions $S_{s\alpha}(\omega)$, $\Delta
P_{s\alpha}(\omega)$, and $\Delta P_{s\alpha}(t)$ can be taken to
characterize the dynamics of the polarization $P_{\alpha}$ in the
effective dynamic theory we want to construct.

Admittedly, the complexity inherent to soft-mode ferroelectrics may
make it difficult to select the feature $i = s\alpha$ in some
cases. Fortunately, a couple of powerful considerations can guide us,
even in the most intricate of situations. On the one hand, recall that
the soft modes in a ferroelectric are likely to present strongly
temperature-dependent frequencies. Hence, as we will see in our
application for PbTiO$_{3}$, inspecting $\bm{S}_{\rm tot}(\omega)$ as
a function of temperature will typically be enough to identify the
features of interest unequivocally. On the other hand, atomistic MD
data can be used to estimate the specific atomic displacements leading
to spectral intensity in specific frequency
ranges.\cite{thomas10,rijal24} Conveniently, such an analysis is
trivial (and of negligible computational cost) in the limit of 0~K,
where all we need to do is calculate and diagonalize the dynamical
matrix corresponding to the ground state of the material. By comparing
the mode eigenvectors thus obtained with the atomic distortions
associated to the spontaneous polarization (readily available from
theory), identifying the key low-frequency oscillators becomes an easy
task.

\subsubsection{Inertial and viscous-damping parameters}

After selecting one mode $i = s\alpha$ for each of the three soft
polar oscillators $P_{\alpha}(t)$, we will typically end up with pairs
of dynamic constants $\{ \gamma_{s\alpha}, \omega_{s\alpha} \}$, as
obtained from fits to the corresponding power spectra
$S_{s\alpha}(\omega)$. Can we use these quantities to derive inertial
($\mu_{\alpha}$) and viscous-damping ($c_{\alpha}$) parameters as
those in Eq.~(\ref{eq:lk})? Note that, in the way we have introduced
the equations of motion (Eq.~(\ref{eq:EoM-i})) and our model for the
power spectrum (Eqs.~(\ref{eq:Si}) and (\ref{eq:cp})), we never deal
with atomic masses or generalized masses associated to the collective
variables $P_{\alpha}(t)$. This is because we proceeded as if we were
working with normal modes, where the massess are used to define a
convenient basis in which oscillators are decoupled both dynamically
and energetically (at the harmonic level). This approach leads
naturally to using $\gamma_{i}$ and $\omega_{i}$ (or the related
quantity $\bar{\omega}_{i}$) as our only fitting parameters, a
situation that is typical in the treatment of experimental spectra
too.

To make contact with Eq.~(\ref{eq:lk}) we proceed as follows. Consider
the equation of motion for $i = s\alpha$ (Eq.~(\ref{eq:EoM-i})) and
let us identify $P_{s\alpha} \equiv P_{\alpha}$. We can multiply all
terms by $\mu_{\alpha}$ to obtain
\begin{equation}
  \mu_{\alpha} \ddot{P}_{\alpha} + c_{\alpha} \dot{P}_{\alpha} +
  \mu_{\alpha}\omega_{\alpha}^{2} \Delta P_{\alpha} = 0 \; ,
  \label{eq:EoM-alpha}
\end{equation}
where we have identified $c_{\alpha} = \mu_{\alpha}\gamma_{\alpha}$
and used $\Delta \dot{P}_{\alpha} = \dot{P}_{\alpha}$. This equation
can be seen as the limit of Eq.~(\ref{eq:lk}) for small
oscillations. Indeed, we can assume
\begin{equation}
  F = F_{eq} + \frac{1}{2} \sum_{\alpha} \mu_{\alpha}
  \omega_{\alpha}^{2} (\Delta P_{\alpha})^{2} + ... \; ,
\end{equation}
where the product $\mu_{\alpha}\omega_{\alpha}^{2} = \kappa_{\alpha}$
is the curvature of the free energy associated to a departure from
equilibrium given by $\Delta P_{\alpha}$.

The proposed connection suggests that, if we were able to obtain an
independent estimate of the curvature $\kappa_{\alpha}$, we could
readily compute $\mu_{\alpha} = \kappa_{\alpha}/\omega_{\alpha}^{2}$
and $c_{\alpha} =
\gamma_{\alpha}\kappa_{\alpha}/\omega_{\alpha}^{2}$. Can we use
atomistic simulations to compute $\kappa_{\alpha}$? This task is in
fact trivial in the limit of 0~K: there, we can determine the atomic
displacements associated to the relevant low-frequency polar phonons
by diagonalizing the dynamical matrix of the ground state; then, by
distorting the ground-state structure following such displacements, we
can compute the energy curvature of interest. By contrast, calculating
$\kappa_{\alpha}$ at finite temperatures is, while technically
possible, significantly less trivial.\cite{thomas10,rijal24}
Nevertheless, we can obtain a reasonable approximation to the
temperature-dependent $\kappa_{\alpha}$ as follows.

The key consideration is that the static dielectric response of the
material is likely to be dominated by the low-frequency polarization
$\bm{P}(t)$ that we want to retain in our time-dependent Landau
theory. More specifically, we can assume
\begin{equation}
  \epsilon_{0} \chi_{\alpha\alpha} = \frac{\partial P_{{\rm
        tot},\alpha}}{\partial \cal{E}_{\alpha}} \approx
  \frac{\partial P_{\alpha}}{\partial\cal{E}_{\alpha}} =
  \kappa_{\alpha}^{-1}\; ,
  \label{eq:chi}
\end{equation}
where $\epsilon_{0}$ is vacuum's dielectric permittivity and we use
the well-known fact that the contribution of an oscillator to the
susceptibility is given by the inverse of the corresponding
free-energy curvature. We thus have
\begin{equation}
  \mu_{\alpha}^{-1} \approx \epsilon_{0} \chi_{\alpha\alpha}
  \omega_{\alpha}^{2} \; .
  \label{eq:mu}
\end{equation}
This is a most convenient relation, because the total susceptibility
tensor $\bm{\chi}$ can be readily obtained from atomistic MD
simulations of the equilibrium state (the corresponding linear
response formulas are derived, e.g., in Ref.~\onlinecite{graf21}).

\subsubsection{An atomistic approach to the polarization mass}
\label{sec:mass}

Ultimately, the dynamics of the soft polarization $\Delta \bm{P}(t)$
involves specific atomic displacements that, as we have just
mentioned, can be readily computed in the limit of 0~K. In the
following we leverage this connection and show how to define an
atom-like mass (given in kg) for the polarization.

Let $\hat{u}_{\alpha,j\beta}$ be a normalized vector specifying the
atomic displacements (of atom $j$ in the unit cell, along direction
$\beta$) contributing to $\Delta P_{\alpha}$. Such
$\hat{u}_{\alpha,j\beta}$ vectors can be identified with the
lowest-lying eigenmodes of the dynamical or force-constant matrix of
the phase of interest. Alternatively, we could derive them from the
atomic distortions associated to the spontaneous polarization itself.

We can compute the polarization change associated to such collective
displacements from knowledge of the Born effective charge tensors
$\bm{Z}_{j}^{*}$. We thus obtain the polarity $p_{\alpha}$,
\begin{equation}
  p_{\alpha} = v^{-1} \sum_{j\beta} Z^{*}_{j,\alpha\beta}
  \hat{u}_{\alpha,j\beta} \; ,
  \label{eq:polarity}
\end{equation}
where $v$ is the unit cell volume. Note that this polarity is
analogous to the quantity used in standard theoretical treatments of
infrared spectra.

We now write $\Delta P_{\alpha}(t) = p_{\alpha}u_{\alpha}(t)$, where
$u_{\alpha}$ quantifies the amplitude of the polarization change and
has units of length. In other words, $u_{\alpha}(t)$ is a dynamical
variable to which we can associate a mass $m_{\alpha}$ with units of
kg. More specifically, by substituting in Eq.~(\ref{eq:lk}), we obtain
the following equation of motion for $u_{\alpha}(t)$
\begin{equation}
  m_{\alpha} \ddot{u}_{\alpha} + c'_{\alpha} \dot{u}_{\alpha} = - v
  \frac{\partial F}{\partial u_{\alpha}} \; ,
  \label{eq:EoM-with-mass}
\end{equation}
where $m_{\alpha} = vp_{\alpha}^{2}\mu_{\alpha}$, $c'_{\alpha} =
vp_{\alpha}^{2}c_{\alpha}$, and the term in the right-hand side is a
force [SI units J~m$^{-1}$]. We have thus introduced a way to compute
an atomic mass $m_{\alpha}$ associated to $\Delta P_{\alpha}(t)$.

Interestingly, knowledge of the vectors $\hat{u}_{\alpha,j\beta}$
allows us to compute a second mass associated to $\Delta
P_{\alpha}(t)$. Specifically, we may consider
\begin{equation}
  \tilde{m}_{\alpha} = \sum_{j\beta} m_{j} \hat{u}_{\alpha,j\beta}^{2}
  \; ,
  \label{eq:mass-tilde}
\end{equation}
where $m_{j}$ is the mass of atom $j$. This kind of effective mass has
been used in effective Hamiltonian studies of polarization dynamics in
ferroelectrics,\cite{ponomareva08,wang11} or whenever a
quantum-mechanical treatment of the polarization was necessary (e.g.,
to investigate quantum paraelectric behavior).\cite{zhong96,iniguez02}
Below we discuss what $m_{\alpha}$ and $\tilde{m}_{\alpha}$ come out
to be, and how they compare with each other, in our example of
PbTiO$_{3}$.

\subsubsection{A comment on units}

Let us briefly note that, as usual, the Landau potential $F$ in
Eq.~(\ref{eq:lk}) is an energy density, with SI units J~m$^{-3}$. The
polarization is an electric dipole per unit volume, thus given in
C~m$^{-2}$. The generalized force acting on the polarization is thus
in J~C$^{-1}$~m$^{-1}$ or V~m$^{-1}$, so it has units of electric
field. The inertial constant $\mu_{\alpha}$ is given in
kg~m$^{3}$~C$^{-2}$, while the viscous damping $c_{\alpha}$ has units
of kg~m$^{3}$~s$^{-1}$~C$^{-2}$. The inertial constant $\mu_{\alpha}$
can be related to an atom-like mass $m_{\alpha}$ (in kg) through a
suitable factor $V p^{2}$, where $V$ is the volume of the considered
supercell (m$^{3}$) and $p$ is a polarity (C~m$^{-3}$) relating atomic
displacements and polarization. Additionally, the modified viscous
damping $c'_{\alpha}$ in Eq.~(\ref{eq:EoM-with-mass}) has units of
kg~s$^{-1}$. Finally, throughout this article we discuss angular
frequencies, $\omega = 2\pi \nu$; for brevity, we indicate the units
of $\omega$ to be THz rather than the actual rad$\times$THz.

\subsection{Numerical application}

We now introduce the material chosen to test and illustrate our
approach to computing dynamic coefficients, ferroelectric perovskite
oxide PbTiO$_{3}$ (PTO for short). We also describe the atomistic
second-principles potential used in our simulations, give the
technical details for the MD runs, and comment on the data processing
to obtain power spectra.

\subsubsection{Ferroelectric perovskite PbTiO$_{3}$}

PbTiO$_{3}$ is a very relevant material, both scientifically and
technologically, and remains one of the best studied
ferroelectrics.\cite{lines-book1977,rabe-book2007} In the present
context, it provides us with a test case that is challenging while
remaining formally simple. The challenge pertains to the non-trivial
lattice-dynamical behavior leading to its ferroelectric phase
transition, which is driven by the softening of a polar phonon
mode. The formal simplicity comes from the fact that the two
structural phases of interest, ferroelectric tetragonal ($P4mm$) and
paraelectric cubic ($Pm\bar{3}m$), present high symmetry and a 5-atom
unit cell, which results in a relatively small number of polar phonons
and, thus, a relatively clean dielectric spectrum.

Experimentally, PTO displays a weakly discontinuous ferroelectric
transition at $T_{\rm C} = 760$~K.\cite{haun87} The spontaneous
polarization is accompanied by a significant lattice deformation, as
the unit cell develops a $c/a$ aspect ratio of about 1.07 at room
temperature, with an elongation along the polar axis. Computational
studies, based on Density Functional Theory
(DFT)\cite{hohenberg64,kohn65} and second-principles
methods,\cite{wojdel13} capture correctly this strong coupling of
polarization and strain.\cite{kingsmith94,wojdel13} Further, in a
recent work we have shown how to derive a third-principles Landau-like
potential that, while being relatively simple, reproduces accurately
the temperature-dependent equilibrium properties of PTO, as regards
both its structural evolution and electromechanical
responses.\cite{pulzone25} It is worth noting here that such a Landau
potential is concerned with the static equilibrium properties of the
compound; in particular, it describes the free-energy landscape for
the total polarization $\bm{P}_{\rm tot}$.

\begin{figure}
    \centering
    \includegraphics[width=0.95\linewidth]{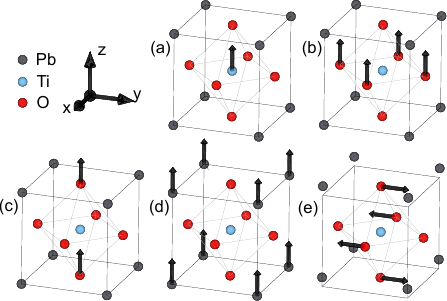}
    \caption{Panels~(a)-(d): Symmetry-adapted atomic distortions of
      the 5-atom cubic perovskite cell that yield a polarization along
      $z$. Panel~(e): In the tetragonal phase (represented here by the
      vertical off-centering of the Pb cations at the cube corners) an
      additional distortion yields a polarization (here along $y$).}
    \label{fig:phonons}
\end{figure}

We can expect the polarization $\bm{P}_{\rm tot}(t)$ of PTO to be
dominated by the polar phonons of the material. These are
$\Gamma$-point (zone-center) modes involving vibrations that occur
homogeneously throughout the lattice, thus respecting the periodicity
set by the 5-atom perovskite unit cell. Such polar phonons can be seen
as linear combinations of symmetry-adapted modes that can be readily
determined for the centrosymmetric cubic phase. We have four such
modes, shown in Figs.~\ref{fig:phonons}(a-d), polarized along the $z$
pseudocubic axis. Because of the cubic symmetry, analogous modes exist
with polarization along $x$ and $y$; they can be obtained from those
in Figs.~\ref{fig:phonons}(a-d) by a suitable rotation of the
structure (e.g., by applying the 3-fold axis about the cell's body
diagonal). The symmetry-adapted modes in Figs.~\ref{fig:phonons}(a-d)
lead to four $z$-polarized phonons at $\Gamma$, one acoustic and three
optical. Hence, in the power spectrum $S_{{\rm tot},z}(\omega)$ of
cubic PTO, we expect to find three phonon-like features (peaks)
corresponding to such optical modes. Then, the cubic symmetry dictates
$S_{{\rm tot},x}(\omega) = S_{{\rm tot},y}(\omega) = S_{{\rm
    tot},z}(\omega)$.

Let us now consider the tetragonal polar phase, assuming, without loss
of generality, that $z$ is the direction of the spontaneous
polarization. In this case, $x$ and $y$ are equivalent directions, but
distinct from $z$, and we expect $S_{{\rm tot},x}(\omega) = S_{{\rm
    tot},y}(\omega) \neq S_{{\rm tot},z}(\omega)$. It can be seen that
the symmetry-adapted modes in Figs.~\ref{fig:phonons}(a-d) continue to
be a complete basis to represent the $z$-polarized phonons, one
acoustic and three optical. Hence, we expect three phonon-like peaks
in $S_{{\rm tot},z}(\omega)$ of tetragonal PTO. For $x$ and $y$,
though, there is an additional way in which the oxygen atoms can
displace creating a net electric dipole. This is sketched in
Fig.~\ref{fig:phonons}(e): because of the spontaneous polar distortion
along $z$ (indicated here by the vertical off-centering of the Pb
cations), the oxygen displacements along $x$, indicated with arrows,
lead to dipoles that need not cancel each other, thus rendering this
distortion polar. As a consequence, the $S_{{\rm tot},x}(\omega) =
S_{{\rm tot},y}(\omega)$ spectra of tetragonal PTO is expected to
present four phonon-like peaks, rather than three.

\subsubsection{Computing $\bm{P}_{\rm tot}(t)$ from atomistic
  simulations}\label{sec:simulations}

To generate the trajectories $\bm{P}_{\rm tot}(t)$, we employ the
second-principles atomistic effective potential for PTO described in
Ref.~\onlinecite{wojdel13}. This model has been used in numerous
studies of PTO and related compounds (e.g., PTO-based superlattices),
and it has been shown to reproduce the behavior of the material in a
way that is qualitatively and quasi-quantitatively
correct.\cite{wojdel14a,zubko16,goncalves19,graf21} The parameters of
our PTO potential were computed by using the local density
approximation (LDA) to DFT.\cite{wojdel13} To compensate for LDA’s
well-known overbinding problem, we simulate the model under the action
of a tensile hydrostatic pressure of
14.9~GPa.\cite{wojdel13,wojdel14b} The only noteworthy discrepancy
between the model predictions and experimental results concerns the
quantification of the ferroelectric transition temperature: for bulk
PTO, the second-principles model yields $T_{\rm C} \approx 510$~K,
while the experimental Curie point occurs at 760~K.\cite{wojdel13}
While significant, this discrepancy is irrelevant in the current
context, as our main concern here is not agreement with
experiment. Rather, our focus is on capturing non-trivial
lattice-dynamical properties using the effective treatment introduced
in Section~\ref{sec:approach}, and our atomistic model for PTO does
provide us with many such challenging behaviors.

To compute the equilibrium trajectory $\bm{P}_{\rm tot}(t)$, we
proceed in the following way. We typically work with a supercell that
is a $10\times 10\times 10$ repetition of the elemental perovskite
5-atom unit, thus corresponding to 5\,000 atoms, assuming periodic
boundary conditions. Close to the computed transition temperature,
between 460~K and 550~K, we increase the supercell size to $12\times
12\times 12$, corresponding to 8\,640 atoms, to mitigate finite-size
effects. At each considered temperature, we first run a Monte Carlo
simulation, following standard procedures discussed
elsewhere,\cite{wojdel13} to determine the equilibrium homogeneous
strain and obtain a characteristic atomic configuration.  We then fix
the homogeneous strain (i.e., we fix the supercell lattice vectors) to
the equilibrium value and, starting from the mentioned typical
equilibrium configuration, run an isokinetic MD simulation with random
initial velocities compliant with Maxwell-Boltzmann's distribution at
the considered temperature. The isokinetic run is 50~ps long, with a
time step of 0.5~fs; it yields a configuration, comprising both atomic
positions and velocities, that we check can be considered as
representative of the equilibrium state. Finally, starting from that
configuration and keeping the homogeneous strains fixed, we run a
production microcanonical, constant-energy and constant-supercell,
simulation for 300~ps with a time step of 0.5~fs, from which we derive
$\bm{P}_{\rm tot}(t)$.  We check these simulation conditions are
sufficient to ensure the energy is conserved and the system fluctuates
around the targeted temperature. (In the results below, the reported
temperatures are obtained from the average kinetic energy of the
microcanonical simulation.)

Thus, in this work we disregard the dynamics of the homogeneous
strains of the system, that is, we keep constant the lattice vectors
defining our simulation supercell. We do this for two reasons. First,
to avoid the difficulties concerning the mass for the lattice vectors
in Parrinello-Rahman dynamics, where making a realistic choice remains
a non-trivial issue in the field of MD simulations at large. Second,
to reduce the complexity of (and avoid spurious features in) the power
spectra we are tasked to fit. Nevertheless, let us stress that
inhomogeneous strains, as corresponding to the long-wavelength
acoustic modulations compatible with our investigated supercell, are
treated exactly in our simulations. Hence, the coupling between local
polarization fluctuations and local strain fluctuations, and all
related effects (e.g., a possible renormalization of the dynamic
constants by strain, or additional anharmonic features in the
spectrum), are accounted for in our treatment.

We compute $\bm{P}_{\rm tot}(\omega)$ from the $\bm{P}_{\rm tot}(t)$
trajectory by a fast Fourier transform (FFT). We find it convenient to
use Welch's method\cite{welch67} to reduce noise in the calculated
$\bm{P}_{\rm tot}(\omega)$, which aids a direct visual consideration
of the spectra while having a negligible impact on the fitted dynamic
parameters. Specificlaly, the time signal was divided into 25~ps
segments with a 50\% overlap, meaning that consecutive segments shared
half of their data. To reduce edge effects, a Hann window was applied
to each segment before calculating its spectrum. The spectra from all
segments were then averaged to obtain the final estimate with a
frequency resolution of about 0.04~THz.

Finally, let us note that $\bm{P}_{\rm tot}$ is computed from the
time-dependent atomic configuration within a linear approximation,
following the standard procedure whereby a centrosymmetric reference
structure is taken to have zero polarization and $\bm{P}_{\rm tot}$ is
obtained from the distortions with respect to such a reference and the
corresponding Born effective charges.\cite{zhong94b,wojdel13}

\section{Results}\label{sec:results}

We first present our computed $\bm{S}_{\rm tot}(\omega)$, to then
discuss our successive approximations to fit its main features.
Finally, we discuss the calculation of the inertial ($\mu_{\alpha}$
and $m_{\alpha}$) and viscous-damping ($c_{\alpha}$ and $c'_{\alpha}$)
constants from the fitted frequencies ($\omega_{\alpha}$) and damping
coefficients ($\gamma_{\alpha}$).

\begin{figure}
    \centering
    \includegraphics[width=0.8\linewidth]{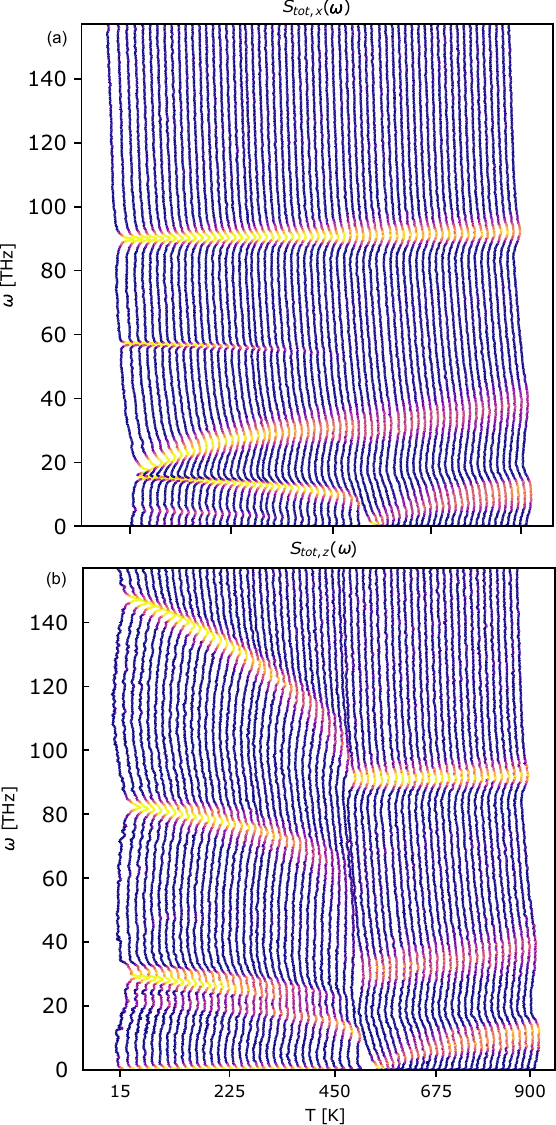}
    \caption{Dielectric power spectrum as obtained for the $x$ (a) and
      $z$ (b) directions. Below $T_{\rm C}$, $z$ is the polar axis. We
      observe $S_{{\rm tot},x}(\omega)=S_{{\rm tot},y}(\omega)$ at all
      temperatures. Then, we have $S_{{\rm tot},x}(\omega)=S_{{\rm
          tot},y}(\omega)=S_{{\rm tot},z}(\omega)$ above the Curie
      point, which occurs at $T_{\rm C}\approx 510$~K. The intensity
      is shown in a linear color scale, where lighter colors
      correspond to a stronger spectral density.}
    \label{fig:Stot_all}
\end{figure}

\subsection{Computed power spectrum}

Figure~\ref{fig:Stot_all} shows the temperature-dependent power
spectra $S_{{\rm tot},x}(\omega)$ and $S_{{\rm tot},z}(\omega)$, in a
linear scale, as obtained from our atomistic simulations following the
procedure described in Section~\ref{sec:simulations}. The data show
some features that shift in frequency strongly as the temperature
changes, including peaks that broaden and tend toward $\omega = 0$ at
$T_{\rm C} \approx 510$~K. We thus reproduce the soft-phonon features
that are known to characterize PTO's ferroelectric
transformation. Note also that $S_{{\rm tot},x}(\omega) = S_{{\rm
    tot},y}(\omega)$ at all temperatures, while $S_{{\rm
    tot},x}(\omega) = S_{{\rm tot},y}(\omega) = S_{{\rm
    tot},z}(\omega)$ above $T_{\rm C}$.

\begin{figure}
    \centering \includegraphics[width=0.95\linewidth]{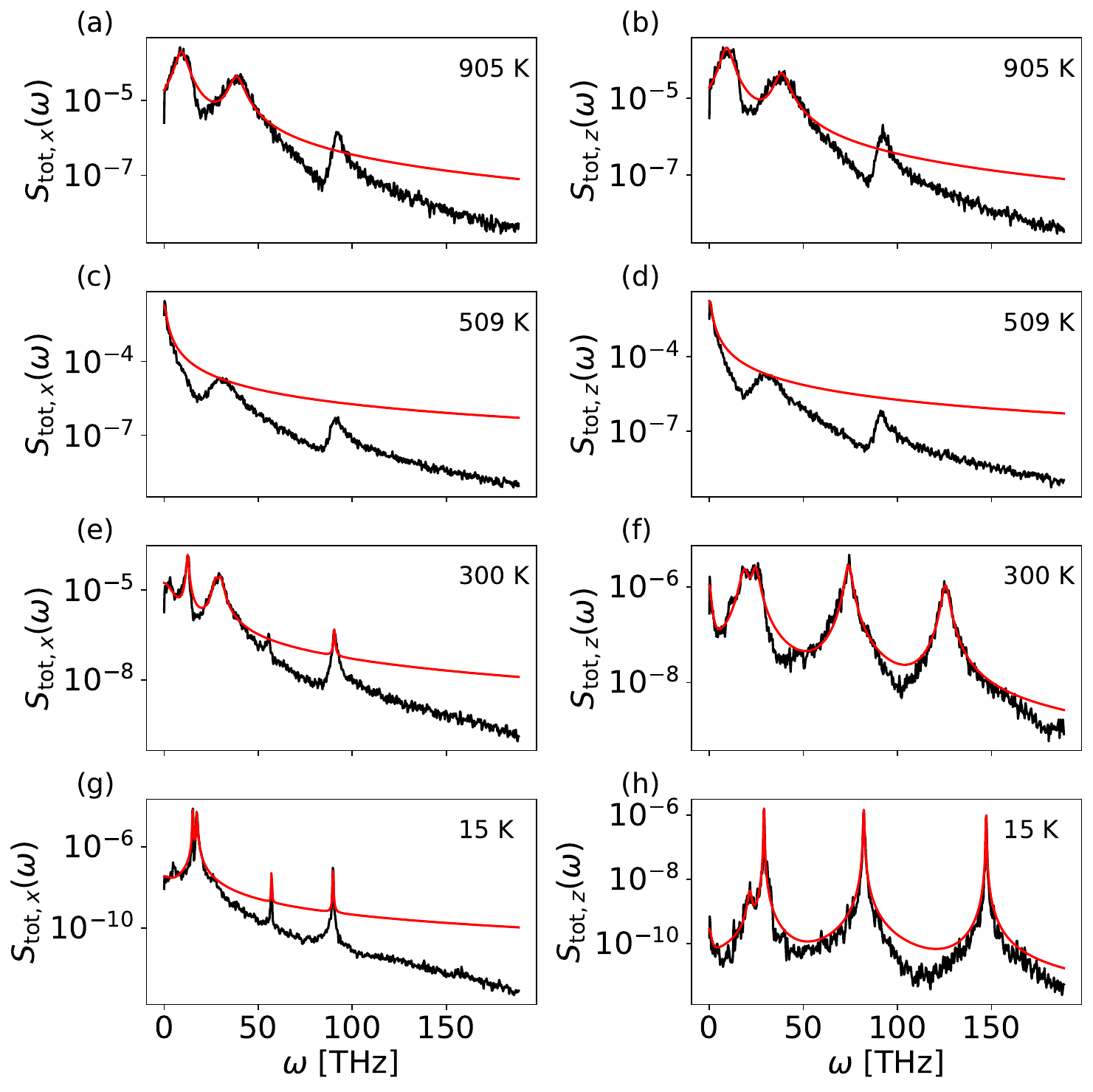}
    \caption{Dielectric power spectra at selected temperatures: 905~K
      in panels~(a) and (b), 509~K in (c) and (d), 300~K in (e) and
      (f), and 15~K in (g) and (h). The black lines are the computed
      spectra. The red lines correspond to fits using four underdamped
      oscillators and one central peak, which we adopt as a sufficient
      choice to treat all temperatures equally (see text). A
      logarithmic scale is used.}
    \label{fig:Stot_some}
\end{figure}

Figure~\ref{fig:Stot_some} shows $S_{{\rm tot},x}$ and $S_{{\rm
    tot},z}$ in a logarithmic scale at selected temperatures, so that
some features can be better appreciated. Well above the computed
$T_{\rm C}$ (905~K, panels (a) and (b)), we find three relatively
broad features, and we observe clearly the expected $S_{{\rm
    tot},x}(\omega) = S_{{\rm tot},z}(\omega)$ symmetry. At $T_{\rm
  C}$ (509~K, panels (c) and (d)), we still see the cubic symmetry,
the spectra having three distinct features, one of them particularly
broad and extending to very low frequencies. Around room temperature
(300~K, panels (e) and (f)), we see a clear difference between the
polar ($z$) and non-polar ($x$) directions. The signatures become
sharper in both spectra, and we find a relatively large number of
distinct features (at least 5 notable peaks for $S_{{\rm tot},x}$, and
at least 6 for $S_{{\rm tot},z}$). Further, we still get a lot of
intensity at very low frequencies. Finally, at very low temperatures
(15~K, panels (g) and (h)) we observe spectra that remain very rich in
features, although one could argue that dominated by a relatively
small number of sharper high-intensity peaks (4 or 5 for $S_{{\rm
    tot},x}$, and 3 strongest ones for $S_{{\rm tot},z}$). Also, the
intensity at low frequencies is relatively modest now.

From inspection of these figures, it is obvious that the expectation
to have a small and symmetry-dictated number of phonon-like peaks in
the spectrum (in the polar phase: 4 for $S_{{\rm tot},x}$ and 3 for
$S_{{\rm tot},z}$) does not apply, as even at the lowest temperatures
we clearly observe more features. Nevertheless, it is possible to
proceed in a systematic way to extract the information pertaining to
the low-frequency polarization $\bm{P}$.

\subsection{Fitting the power spectrum}

Let us describe our procedure by considering cases of increasing
complexity.

At 905~K, well above $T_{\rm C}$, the spectrum is composed of three
broad but distinct peaks, which we can identify with the three polar
phonon modes that PTO is expected to present in its cubic phase. As
shown in Fig.~\ref{fig:Stot_some}, we can easily fit the two
lowest-frequency peaks. As for the third peak, whose intensity is more
than $100\times$ smaller than that of the strongest one, it is all but
impossible to capture it unless we carefully initialize and constrain
the fitting parameters. (Since such high-frequency features are not
our concern here, we do not pursue this issue any further.)  For the
lowest-frequency peak at 905~K, we obtain $\omega_{sz} = 9.7$~THz and
$\gamma_{sz} = 5.9$~THz, which is consistent with the underdamped
oscillator assumption. Thus, from the analysis of our data at 905~K,
we conclude that a model spectrum containing two (or three)
underdamped oscillators allows us to obtain reliable results in the
frequency range of interest. We find that this situation is typical of
the paraelectric phase, extending to temperatures as low as $T =
556$~K, where we can still fit an underdamped soft mode reliably.

Let us now consider the data at 15~K, that is, in the limit of low
temperatures. The $S_{{\rm tot},z}$ spectrum, corresponding to the
direction of the spontaneous polarization, seems particularly
clear. It displays three intense and sharp peaks that can be readily
fitted by considering three underdamped oscillators in the model
spectral function. Then, a fourth feature at about 21.5~THz can also
be readily fitted by introducing a fourth oscillator in our model
function. In all cases, the fitted frequencies and damping constants
are compatible with the underdamped oscillator assumption.

The case of the $S_{{\rm tot},x}$ spectrum at 15~K is more involved,
as the number of sharply defined features is considerably
larger. Here, as shown by the red line in Fig.~\ref{fig:Stot_some}(g),
we can obtain accurate fits of the main features in the spectrum by
including four underdamped oscillators and a central peak. But then,
while fitting the $S_{{\rm tot},\alpha}(\omega)$ spectra at 15~K is
relatively straightforward, a question obviously emerges: how shall we
select the feature corresponding to the low-frequency polarization
$\bm{P}(t)$ that we want to treat in our dynamical Landau theory? This
critical issue will be discussed in the following section.

Finally, the results at 300~K are representative of a broad interval
of temperatures (from about 200~K to about 400~K) where the computed
spectra are most challenging to fit, for two main reasons. First, for
both $S_{{\rm tot},x}$ and $S_{{\rm tot},z}$ we find significant
intensity at low frequencies. We check that such features are
incompatible with the underdamped oscillator assumption, and thus
treat them as a central peak (Eq.~(\ref{eq:cp})). Second, in $S_{{\rm
    tot},z}$ we find two broad peaks at about 23~THz that need to be
resolved in detail, as both could in principle correspond to the
low-frequency polarization we want to model in our dynamical Landau
theory. Despite these difficulties, as shown in
Figs.~\ref{fig:Stot_some}(e) and \ref{fig:Stot_some}(f), by
considering a model spectrum that includes four underdamped
oscillators together with a central peak contribution, all relevant
features can be captured in a satisfactory way without any need for
careful adjustments of the fitting procedure. Indeed, this model
spectral function (i.e., four oscillators and a central peak) can be
safely applied to fit the computed $\bm{S}_{\rm tot}(\omega)$ at all
temperatures, yielding consistent results for the frequencies and
damping constants of the features of interest. All the results
presented in the following correspond to this choice of model spectral
function, which leads to the fits depicted with red lines in
Fig.~\ref{fig:Stot_some}.

\subsection{Identifying the relevant spectral features}\label{sec:identify}

We now show how the calculation of the polar phonon modes and
dielectric spectra in the limit of 0~K provides us with a guideline to
interpret the features observed in $\bm{S}_{\rm tot}(\omega)$ at
finite temperatures.

\setlength{\extrarowheight}{1pt}
\setlength{\tabcolsep}{8pt}
\begin{table}[h!]
\caption{Zone-center phonon frequencies (in THz) for the tetragonal
  ferroelectric phase of PbTiO$_{3}$ as obtained by diagonalizing the
  dynamical matrix $\bm{\mathcal{D}}$ of the ground state (column
  denoted by $\bm{\mathcal{D}}$(0~K)) and by computing the dielectric
  power spectrum from an MD simulation at 1~K for a 5-atom simulation
  cell (${\bm S}_{\Gamma}$(1~K)). We indicate whether the modes are
  polar or non-polar, and whether the atomic displacements are along
  the $z$ polar axis or in the $xy$ plane. In the latter case, the
  frequencies correspond to degenerate mode pairs.}
\vskip 2mm
\centering
\begin{tabular}{ccc}
\hline\hline
$\bm{\mathcal{D}}$(0~K) & ${\bm S}_{\Gamma}$(1~K) & Character \\
\hline
15.0 & 15.0 & $xy$ polar soft \\
16.0 & 16.0 & $xy$ polar \\
29.3 & 29.3 & $z$ polar soft  \\
57.2 & 57.2 & $xy$ polar \\
65.3 & --   & $z$ non-polar  \\
82.5 & 82.5 & $z$ polar  \\
89.7 & 89.7 & $xy$ polar \\
148.2 & 148.2 & $z$ polar \\
\hline\hline
\end{tabular}
\label{tab:freq}
\end{table}

Table~\ref{tab:freq} lists the frequencies of the $\Gamma$-point
phonons as computed for the $P4mm$ ground state of PTO. Here we use
our second-principles model to represent the potential energy surface,
and apply a standard procedure to obtain the dynamical matrix
($\bm{\mathcal{D}}$(0~K)) numerically in an essentially exact
way.\cite{phonopy} The Table also describes the character of the
different modes, as determined from inspection of the corresponding
eigenvectors. We readily identify the soft modes $i = s\alpha$, whose
eigenvectors resemble closely the atomic distortions associated to the
spontaneous polarization. They are the lowest-frequency phonons,
confirming that -- at this level at least -- PTO behaves in the
expected simple manner. Note that, in this strictly harmonic limit, we
recover the number of polar modes expected by symmetry, i.e., four in
the $x \equiv y$ directions and three polarized along $z$.

\begin{figure}
  \centering \includegraphics[width=\linewidth]{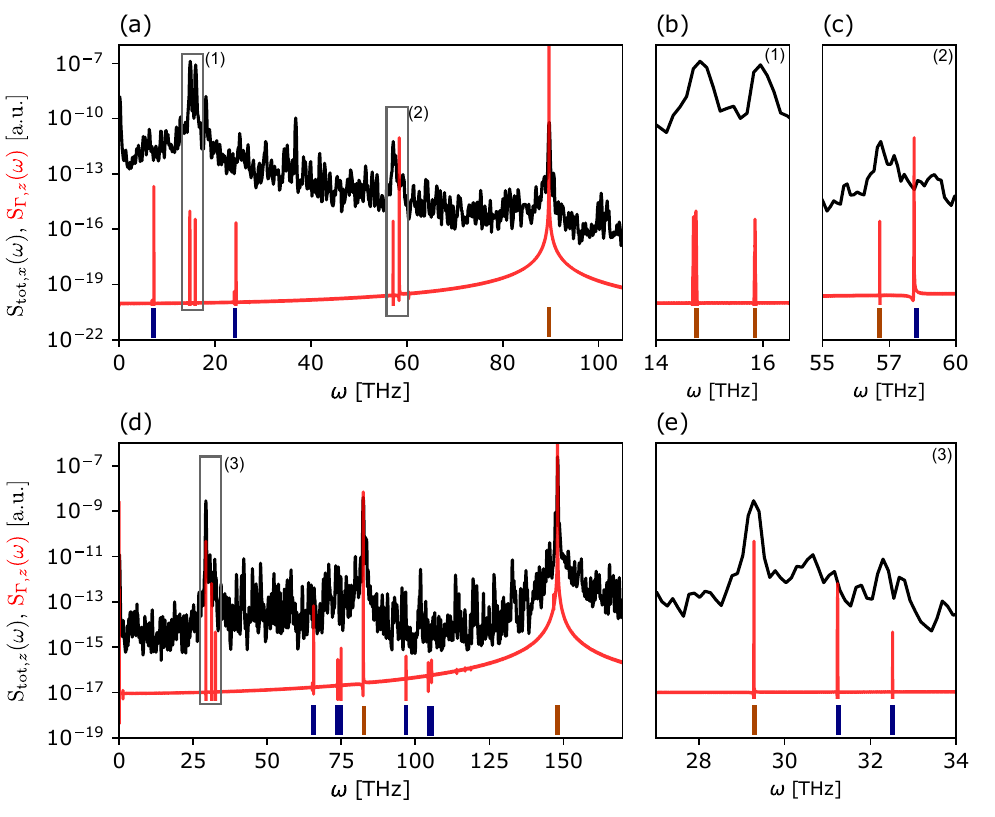}
\caption{Dielectric power spectra of tetragonal PTO computed at
  1~K. Panels (b), (c), and (e) show enlarged views of the numbered
  boxes in panels (a) and (d).  Red lines show the spectra obtained
  from an MD simulation of a $1\times 1\times 1$ supercell, denoted as
  $\bm{S}_{\Gamma}(\omega)$. Black lines show the spectra computed at
  1~K, but now from an MD simulation of a $10\times 10\times 10$
  supercell. The additional oscillators introduce noise and anharmonic
  features at low intensities. Nevertheless, the same dominant phonon
  peaks (marked by short dark red vertical lines) can be identified in
  both spectra; the corresponding frequencies match those listed in
  Table~\ref{tab:freq} and obtained from a direct diagonalization of
  the dynamical matrix of the tetragonal ground state
  ($\bm{\mathcal{D}}$(0~K)). The short blue lines mark overtones,
  i.e., anharmonic features not present in the phonon list of
  Table~\ref{tab:freq}. For example, the $S_{\Gamma,x}$ spectrum (a)
  displays a strong overtone at about 7.2~THz, which can be seen as
  originating from a third-order coupling that involves the
  $x$-polarized phonon at 89.7~THz (quadratically) and the
  $z$-polarized phonon at 82.5~THz (linearly), such that
  7.2~THz~$\approx$~89.7~THz~$-$~82.5~THz. Such an overtone disappears
  when we restrict the motion of the atoms, in the MD simulation, to
  be only along $x$, supporting our interpretation that it comes from
  a crossed coupling of the form $\sim x^{2}z$. As a second example,
  the overtone at about 96.9~THz in the $S_{\Gamma,z}$ spectrum (d) is
  most likely reflecting a biquadratic coupling of the form $\sim
  x^{2}z^{2}$, involving the same two phonons, which leads to
  96.9~THz~$\approx 2\times 89.7$~THz~$-$~82.5~THz. Similar low-order
  phonon-phonon anharmonic couplings (and the corresponding frequency
  combinations) can be identified to account for all the overtones
  observed.}
\label{fig:1K}
\end{figure}

Table~\ref{tab:freq} also lists the frequencies obtained from an MD
simulation at 1~K, performed following the procedure described in
Section~\ref{sec:approach}, but considering a $1\times 1\times 1$
simulation cell. Such MD run allows us to explore the vicinity of the
ground state in a perturbative manner, restricted to $\Gamma$-point
distortions. The corresponding spectral function $\bm{S}_{\Gamma}$,
shown in Fig.~\ref{fig:1K} (in red), displays well-defined
features. In particular, we find peaks for all the polar phonons
computed exactly in the limit of 0~K and listed in
Table~\ref{tab:freq}. These features are marked with short red
vertical lines in the figure.

Figure~\ref{fig:1K} shows that the computed 1~K spectra display
additional peaks (marked with short blue vertical lines) that do not
correspond to eigenvectors of $\bm{\mathcal{D}}$(0~K). Such features
can thus be identified as overtones emerging from anharmonic
phonon-phonon couplings. Indeed, it can be seen that their frequencies
correspond to simple combinations of the phonon frequencies (see
examples in the figure caption). It is interesting to note that, in
some cases, the anharmonic couplings involve phonons polarized along
different pseudocubic directions (see, e.g., the peak at 7.2~THz in
the $S_{\Gamma,x}$ spectrum); we can unequivocally establish this by
running MD simulations in which vibrations along particular directions
are precluded, which results in the disappearance of the feature.

Figure~\ref{fig:1K} also shows (in black) the spectra obtained from
1~K MD but considering the $10\times 10\times 10$ supercell: the
spectra are much noisier in this case, but the dominant phonon peaks
remain clear and they match those observed in $\bm{S}_{\Gamma}$.

The exercise just described allows us to identify the
$P_{s\alpha}(\omega)$ features of interest in the limit of low
temperatures. Then, by following such spectral features as the system
heats up, we can work out their temperature evolution and make contact
with the spectra of Fig.~\ref{fig:Stot_all}. Of special note is the
case of the two close peaks found in $S_{{\rm tot},z}$ and mentioned
above when discussing Figs.~\ref{fig:Stot_all}(b) and
\ref{fig:Stot_some}(f). We find that only one of them has a
significant intensity at very low temperatures: it clearly corresponds
to the $z$-polarized soft phonon at 29.3~THz in the limit of 0~K. The
other peak becomes significant only as we heat up, and it seems to
correspond to an overtone whose precise nature is difficult to
pinpoint. We discuss it below.

\begin{figure}[t!]
    \centering
    \includegraphics[width=0.8\linewidth]{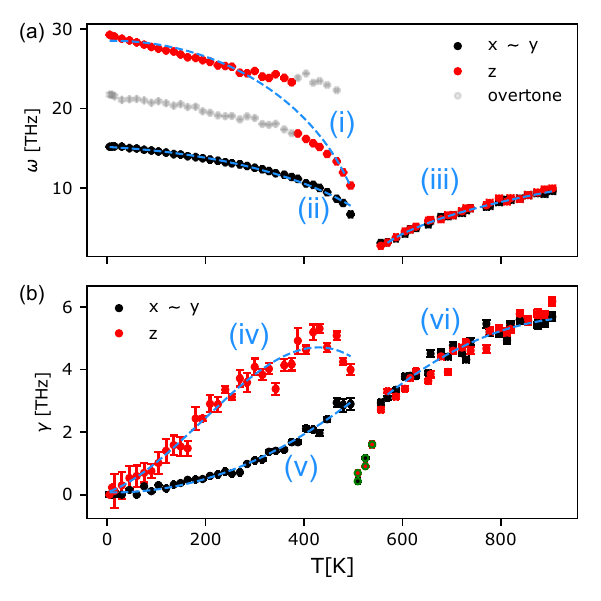}
    \caption{Fitted frequencies (a) and damping constants (b) for
      $P_{z}$ (red) and $P_{x} \sim P_{y}$ (black). Gray symbols in
      panel~(a) correspond to an overtone that anticrosses with
      $P_{z}$. Dashed blue lines are polynomial fits specified in
      Table~\ref{tab:fits}. Between 509~K and 539~K, approximately,
      the soft oscillators become overdamped; hence, in that range,
      panel~(a) shows no frequencies and panel~(b) displays (outlined
      in green) the damping constants corresponding to the fitted
      central peaks.}
    \label{fig:fits}
\end{figure}

\begin{table}[t!]
\centering
\caption{Polynomial fitting functions shown in Figs.~\ref{fig:fits}
  and \ref{fig:mass}. For these fits, we use a reduced temperature
  $\bar{T}=T/T_{\rm C}$, with $T_{\rm C} = 510$~K. The units are
  implicit; we use THz$^{2}$ for $\omega^{2}_{\alpha}$, THz for
  $\gamma_{\alpha}$, and kg~s$^{-1}$ for $c'_{\alpha}$, while
  $m_{\alpha}/m_{\rm Pb}$ is dimensionless.}
\vskip 2mm
\centering
\begin{tabular}{cl}
  \hline\hline Label & Fitting function \\ \hline\rule{0pt}{2.5ex}
(i) & $\omega_{z}^{2} = 813.11 - 27.59\bar{T} -783.07\bar{T}^{2}$ \\ 
(ii) & $\omega_{x}^{2} = 229.71 -49.94\bar{T} -128.64\bar{T}^{2}$ \\ 
(iii) & $\omega_{x}^{2} \sim \omega_{z}^{2} = -165.21 + 176.24\bar{T} -17.72\bar{T}^{2}$ \\ 
(iv) & $\gamma_{z} = 0.07 +  2.78\bar{T} + 12.95\bar{T}^{2} - -11.55\bar{T}^3$ \\ 
(v) & $\gamma_{x} = 0.07 - 0.16\bar{T} + 3.21\bar{T}^{2}$ \\
(vi) & $\gamma_{x}^{2} \sim \gamma_{z}^{2}= -9.00 + 15.47\bar{T} - 4.09\bar{T}^{2}$ \\ 
(vii) & $m_{z}/m_{\mathrm{Pb}} = 0.94 - 0.58\bar{T} + 0.34\bar{T}^{2}$ \\ 
(viii) & $m_{x}/m_{\mathrm{Pb}} = 0.14 + 0.37\bar{T} - 0.21\bar{T}^{2}$ \\ 
(ix) & $m_{x}/m_{\mathrm{Pb}} \sim m_{z}/m_{\mathrm{Pb}} = 2.30 - 2.19\bar{T} + 0.69\bar{T}^{2}$ \\ 
(x) & $c'_{z} = 10^{-12}\left(-0.006 + 0.23\bar{T} - 0.10\bar{T}^{2}\right)$ \\ 
(xi) & $c'_{x} = 10^{-12}\left(0.0007 - 0.005\bar{T} + 0.04\bar{T}^{2}\right)$ \\ 
(xii) & $c'_{x} \sim c'_{z} = 10^{-12}\left(-0.008+ 0.10\bar{T} - 0.015\bar{T}^{2}\right)$ \\[0.5mm]
\hline\hline
\end{tabular}
\label{tab:fits}
\end{table}

\subsection{Effective frequencies and damping constants}

Figure~\ref{fig:fits} shows the frequencies $\omega_{\alpha}$ and
damping constants $\gamma_{\alpha}$ corresponding to the soft
polarization, as a function of temperature, following the assignment
of spectral features described above. Such a temperature dependence
can be fitted using simple polynomial functions, as given in
Table~\ref{tab:fits}. Note that near $T_{\rm C}$ the soft mode becomes
overdamped (in the fit, the corresponding intensity becomes part of
the central peak); as a result, we do not report any frequencies
there, but we do include the central-peak damping constants in
Fig.~\ref{fig:fits}(b).

\begin{figure}[t!]
    \centering
    \includegraphics[width=0.8\linewidth]{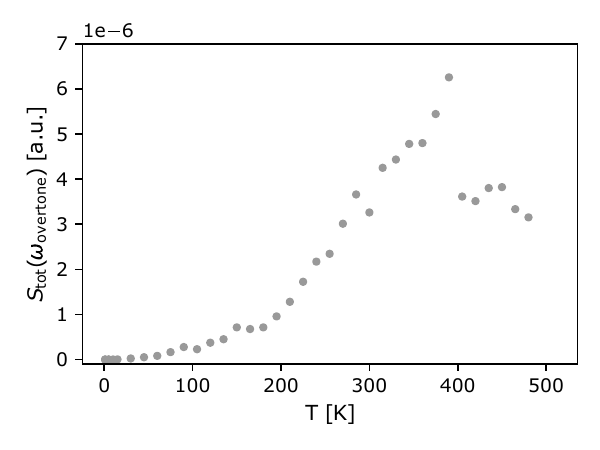}
    \caption{Fitted spectral intensity of the overtone peak whose
      frequency is shown in grey in Fig.~\ref{fig:fits}(a), as a
      function of temperature.}
    \label{fig:overtone}
\end{figure}

Figure~\ref{fig:fits}(a) also shows (in grey) the frequency of the
mentioned overtone, which anticrosses with our soft oscillator of
interest $P_{z}(\omega)$. Finding the origin of this overtone was a
focus of attention for us. We ran detailed calculations in the limit
of $T\rightarrow 0$~K (not shown here) to try to identify a clean
spectral feature as those in Fig.~\ref{fig:1K}. While confirming the
presence of the overtone, the calculations also revealed that its
intensity decreases as we go down in temperature, as shown in
Fig.~\ref{fig:overtone}. Indeed, the intensity is so small as we
approach 0~K that this feature is impossible to detect in the spectra
of Fig.~\ref{fig:1K}. Hence, we conclude that this feature requires a
sizable phonon population for it to be significant, which suggests it
is associated to an anharmonic coupling of relatively high order,
potentially involving hard (and maybe non-$\Gamma$) phonons. An
assignment would require the use of costly techniques to resolve the
atomic motions associated to certain frequency
ranges,\cite{thomas10,rijal24} and we have not attempted it here.

Beyond that intriguing feature, the polarization $\bf{P}(\omega)$
behaves in the expected manner. As regards the characteristic
frequencies, they are always in the range of tens of THz. Further, we
obtain a behavior that matches our expectations for a
non-reconstructive phase transition driven by soft modes, with
$\omega_{\alpha}$ approaching zero from above and below $T_{\rm C}$
with an approximate $|T-T_{\rm C}|^{1/2}$ dependence. We also recover
the expected symmetries ($\omega_{x} = \omega_{y} = \omega_{z}$ for
$T>T_{\rm C}$ and $\omega_{x} = \omega_{y} \neq \omega_{z}$ for
$T<T_{\rm C}$) as well as the expected stronger hardening of the polar
phonon that condenses in the phase transition
($\omega_{z}>\omega_{x}=\omega_{y}$ below $T_{\rm C}$).

As for the damping constants $\bm{\gamma}$, we obtain values of a few
THz at most, indicating relaxation times of a fraction of ps. The
obtained damping constants increase gradually with temperature,
reflecting the growing anharmonicity and shortening equilibration
times of the system upon heating. We find that $\gamma_{z}$ departs
from this smooth behavior in a wide temperature range below $T_{\rm
  C}$, suggesting a faster relaxation for the $\Delta P_{z}$ component
(up to three to four times faster). The relatively large $\gamma_{z}$
values are obtained in the temperature range where the mentioned
anticrossing with an overtone occurs; this obviously impacts the shape
of the $P_{z}(\omega)$ peak and the corresponding damping constant,
suggesting the possibility that the observed $\gamma_{z} >
\gamma_{x}=\gamma_{y}$ might be partly coincidental and dominated by
the coupling of $P_{z}$ with the overtone. Alternatively, it makes
good physical sense for the soft oscillator associated to the
spontaneous polarization to present relatively strong anharmonic
couplings to other phonons (i.e., to the thermal bath), on account of
the distortion associated to $P_{z,eq}\neq 0$ below $T_{\rm C}$. At
any rate, our second-principles results do suggest an enhanced damping
for the $\Delta P_{z}$ component, whatever its origin.

\begin{figure}
    \centering \includegraphics[width=0.8\linewidth]{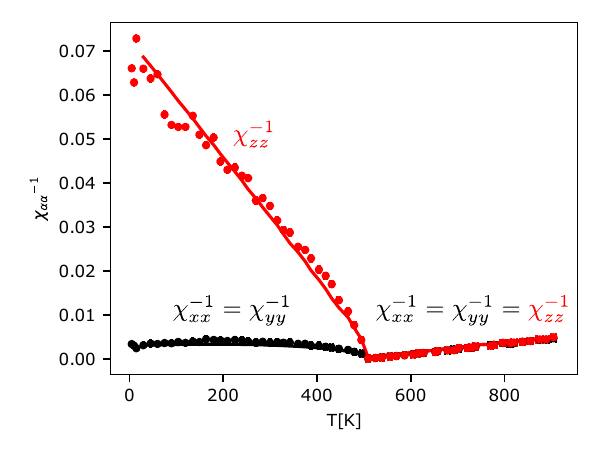}
   \caption{Temperature dependence of the inverse dielectric
     susceptibility $\chi_{\alpha\alpha}^{-1}$
     (dimensionless). Symbols correspond to the results obtained from
     MD simulations directly; solid lines correspond to the results
     derived from the Landau model labeled $F_{9}$ in
     Ref.~\onlinecite{pulzone25}. Red and black denote the $zz$ and
     $xx = yy$ components, respectively.}
    \label{fig:chi}
\end{figure}

\subsection{Inertial and viscous-damping constants}

As explained in Section~\ref{sec:method}, we can compute inertial
constants $\mu_{\alpha}$ for the polarization components $\Delta
P_{\alpha}(t)$ provided we have access to the curvature
$\kappa_{\alpha}$ of the free energy landscape around the equilibrium
state. Conveniently, $\kappa_{\alpha}$ can be obtained from the static
susceptibility $\epsilon_{0}\chi_{\alpha\alpha}$ (Eq.(\ref{eq:chi})),
assuming that $\Delta \bm{P}_{tot} \approx \Delta \bm{P}$ is a good
approximation for the purpose of computing the static dielectric
response of the system. Further, as described in
Section~\ref{sec:mass}, we can connect the inertial constants to
atom-like masses $m_{\alpha}$ for the polarization
(Eq.~(\ref{eq:polarity})). To do that, we estimate the polarities
$p_{\alpha}$ by identifying the $\hat{u}_{\alpha,j\beta}$ vectors with
the soft eigenvectors of the force-constant matrix at 0~K.

\begin{figure}
    \centering
    \includegraphics[width=0.8\linewidth]{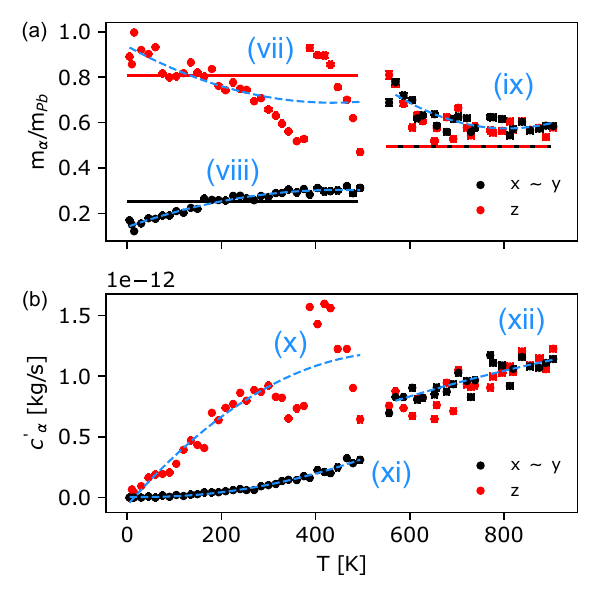}
    \caption{Panel~(a): Temperature dependence of the normalized mass
      ratio, $m_{\alpha}/m_{\mathrm{Pb}}$, for $P_{z}$ (red) and
      $P_{x} \sim P_{y}$ (black). Panel~(b): Viscous-friction
      constants $c'_{\alpha}$. The blue dashed lines are polynomial
      fits, and the labels (vii)--(xii) correspond to the fitting
      functions given in Table~\ref{tab:fits}. The masses $m_{\alpha}$
      correspond to those of Eq.~(\ref{eq:EoM-with-mass}), and they
      can be related to the inertial constants $\mu_{\alpha}$ in
      Eq.~(\ref{eq:EoM-alpha}) by using the cell volume
      $v=58.34$~\AA$^{3}$ and polarity $p=2.46\times
      10^{-20}$~C~\AA$^{-3}$, which can be taken as temperature
      independent for this purpose. Similarly, the damping constants
      $c_{\alpha}$ can be derived from the $c'_{\alpha}$ of
      panel~(b). The solid horizontal lines in panel~(a) correspond to
      the effective polarization masses obtained using
      Eq.~(\ref{eq:mass-tilde}) (see text).}
    \label{fig:mass}
\end{figure}

We thus proceed by computing the susceptibility tensor $\bm{\chi}$
from our MD simulations by using a well-known result from
linear-response theory.\cite{graf21} (Alternatively, we can calculate
the tensor from the Landau model for PTO that we have recently
reported\cite{pulzone25} and which was computed from a training set of
second-principles data generated with the same effective potential
used in this work.) Figure~\ref{fig:chi} shows the results thus
obtained for $\bm{\chi}$. From the data in this figure and in
Fig.~\ref{fig:fits}, we evaluate Eq.~(\ref{eq:mu}) to obtain the
temperature-dependent inertial constants $\mu_{\alpha}$. And from
those we readily derive the temperature-dependent masses $m_{\alpha}$
shown in Fig.~\ref{fig:mass}(a).

Note that the results for $m_{z}$ around 400~K indicate that this mass
inherits the discontinuity in the corresponding frequency
$\omega_{z}$, caused by the mentioned overtone anticrossing
(Fig.~\ref{fig:fits}). (An equivalent discontinuity is obtained for
$\mu_{z}$, not shown here.) A word of caution is in order here: By
using the total susceptibility $\bm{\chi}$ to estimate $\kappa_{z}$,
we assume a smooth behavior for this quantity, so that the
anticrossing in $\omega_{z}$ is fully absorbed by $m_{z}$. In reality,
there are physical reasons to expect small discontinuities or
anomalies in both the free-energy curvature $\kappa_{z}$ and the mass
$m_{z}$, since the anticrossing with the overtone will impact both the
atomic displacements associated to $P_{z}$ (and thus its effective
mass) as well as the associated free-energy landscape. However, noting
that the effects are relatively small, and that clarifying this issue
would be non-trivial, we do not pursue it further.

Figure~\ref{fig:mass}(a) also shows, as horizontal lines, the direct
estimate of the polarization masses, $\tilde{m}_{\alpha}$, from the
$\hat{u}_{\alpha,j\beta}$ atomic displacements using
Eq.~(\ref{eq:mass-tilde}). To compute the $\tilde{m}_{\alpha}$ masses,
we consider the force-constant matrix for the tetragonal ground state
as well as the one corresponding to the cubic state that is a saddle
point of the energy at 0~K. The corresponding soft-mode eigenvectors
allow us to estimate $\tilde{m}_{x}=\tilde{m}_{x}\neq \tilde{m}_{z}$
for the ferroelectric phase, as well as
$\tilde{m}_{x}=\tilde{m}_{x}=\tilde{m}_{z}$ for the paraelectric
phase. We find a remarkable agreement between $m_{\alpha}$ and
$\tilde{m}_{\alpha}$, obtaining values that are a fraction of the mass
of the Pb atom. This makes good physical sense given that
ferroelectricity in PTO is largely driven by Pb's lone-pair chemistry
and the formation of Pb--O bonds. Further, the approximate agreement
between $m_{\alpha}$ and $\tilde{m}_{\alpha}$ implies that the
obtained mass anisotropies (with $m_{x}=m_{y} \approx 0.25\, m_{z}$ in
the ferroelectric phase) are essentially related to differences in the
corresponding soft-mode eigenvectors.

Finally, note that we can also compute the viscous-friction constants
$c_{\alpha}$ in Eq.~(\ref{eq:lk}), as well as the related
$c'_{\alpha}$. The corresponding results are given in
Fig.~\ref{fig:mass}(b).

\section{Discussion}\label{sec:discussion}

We now comment on a number of aspects concerning our results for PTO
and the general methodology to derive effective dynamic constants for
simulations at the mesoscopic or macroscopic scales.

\subsection{Comparison with experiment}

Our results indicate a monotonic and smooth increase of the damping
constants -- $\bm{\gamma}$ or, equivalently, $\bm{c}'$ -- with growing
temperature. Except for the mentioned anomaly in $\gamma_{z}$ below
$T_{\rm C}$, probably caused by the (accidental) anticrossing with an
overtone, our results for $\bm{\gamma}$ comply with general
expectations. Indeed, when phonon lifetimes are controlled by
phonon-phonon scattering, as in the case of our atomistic MD
simulations, we expect to have $\gamma_{\alpha}$ proportional to
phonon population. Then, in our classical MD runs, phonon population
is essentially proportional to $T$, which should lead to $\bm{\gamma}$
approximately linear with temperature (at least, in the limit of a
perfectly harmonic crystal), which agrees with the general trend we
indeed find.

When trying to connect with experimental results, one must note that
phonon-phonon scattering at low temperatures will be essentially zero
(on account of quantum statistics) and damping will be dominated by
other forms of scattering, e.g., by defects. We do not have any of
those effects in our MD simulations and, thus, a quantitative
comparison at low temperature is not warranted. By contrast, at
high-$T$ the Bose-Einstein occupation becomes proportional to $T$ and
we expect $\gamma_{\alpha} \propto T$ dominated by phonon-phonon
scattering, as in our MD simulations,\cite{kittel-book1966} which
should make a quantitative comparison viable. We should nevertheless
bear in mind the discrepancy between our predicted $T_{\rm C}$ (510~K)
and the experimental one (760~K).

Having said this: Fontana {\sl et al.}\cite{fontana91} reported
frequencies for PTO's soft polar modes in the ferroelectric
phase. Their results are compatible with our findings, as they observe
the same sort of soft-mode behavior and $T$-dependent frequencies that
we obtain. Further, at 273~K (their lowest reported temperature) they
get about 28~THz for $P_{z}$ (labeled $A_{1}$ in
Ref.~\onlinecite{fontana91} on account of the phonon's symmetry) and
about 17~THz for $P_{x} \sim P_{y}$ (labeled $E$ in
Ref.~\onlinecite{fontana91}). For comparison, in a comparable $T$
range (Fig.~\ref{fig:fits}(a)) we get about 25~THz and 15~THz,
respectively. These authors also report $\gamma_{x}=\gamma_{y}$,
corresponding to the $E$ modes. They see a sustained increase when
approaching $T_{\rm C}$ from below, with $\gamma_{x}$ reaching values
of about 4~THz. Both the general behavior and the quantitative values
(we get 3~THz; see black symbols as $T\rightarrow T_{\rm C}$ from
below in Fig.~\ref{fig:fits}(b)) are compatible with our results.

Hlinka {\sl et al.} reported related data, but this time above $T_{\rm
  C}$.\cite{hlinka13} They observe a soft mode behavior compatible
with our results. Specifically, at about 100~K above $T_{\rm C}$ they
measure a frequency of about 1~THz, while we get frequencies of about
3~THz in the same range. As for the damping, they obtain values around
1~THz for $T-T_{\rm C} \approx 100$~K, while we get about 3~THz in the
corresponding temperature range. More interestingly, the soft mode
becomes overdamped about 50~K above $T_{\rm C}$ in the experiment,
which agrees well with our observation: in Fig.~\ref{fig:fits}(a), no
results are shown in the range from 509~K to 539~K on account of our
soft mode being overdamped. Hence, we conclude our results are
generally consistent with the experiments of
Ref.~\onlinecite{hlinka13}. (When comparing our results and those in
Ref.~\onlinecite{hlinka13}, there is a subtlety regarding the
criterion to consider a mode as being overdamped, which depends on
whether one is modeling real time dynamics -- as done here -- or a
particular spectral function directly.)

To conclude this part, we should note that Foster {\sl et al}.
reported on anharmonic effects affecting the line shape of the $A_{1}$
mode (i.e., our $P_{z}$) in a range of temperatures between 225~K and
400~K.\cite{foster93a} This is strongly reminiscent of the overtone we
find in the $S_{{\rm tot},z}$ spectrum, in essentially the same range
of frequencies as the soft mode $P_{z}$. We lack detailed information
to explore this connection further. Nevertheless, it is certainly
intriguing and encouraging that both experiment and simulation reveal
a distinctive feature affecting, specifically, the soft oscillator
corresponding to PTO's spontaneous polarization.

\subsection{Implications for dynamic simulations based on
  non-atomistic effective theories}

A growing number of articles discuss Ginzburg-Landau phase-field
simulations where the polarization field $\bm{P}(\bm{r},t)$ evolves
according to an equation of motion equivalent to Eq.~(\ref{eq:lk}),
considering explicitly both inertial and damping
terms.\cite{akamatsu18,yang20,li21b,chen24,wang25} Typically, such
studies assign values to the corresponding dynamic constants by
fitting relevant experimental results. Further, as far as we can tell,
all existing works make some implicit assumptions, most notably, that
the tensors $\bm{\mu}$ and $\bm{\gamma}$ are isotropic and do not
depend on temperature or the state of the system. Similar
approximations are implicit in MD studies based on effective
Hamiltonians of the kind originally introduced by Rabe, Vanderbilt and
others:\cite{zhong94a,zhong95a,ponomareva08,wang11,zhong96,iniguez02}
temperature-independent masses are assigned to local electric dipoles
(computed in essentially the same way as done here to derive
$\tilde{\bm{m}}$ in Section~\ref{sec:mass}) and the equilibration is
assumed to occur in a time scale that is state and temperature
independent (as dictated by the friction coefficient in Langevin's
dynamics, for example).

Our results shed light on the validity of such unchecked
approximations. (PTO being a prototype compound, we suspect our main
conclusions will be relevant to all soft-mode ferroelectrics,
certainly perovskites.) For example, we find that the masses
$m_{\alpha}$ (equivalently, the inertial constants $\mu_{\alpha}$) are
state dependent. Most remarkably, they change significantly (by about
a factor of 2) across the ferroelectric phase transition. Further, in
the ferroelectric phase the masses reflect the crystallographic
symmetry breaking, as polarization changes parallel to the polar axis
experience an inertia that is three to four times bigger than that
governing polar vibrations perpendicular to it (i.e., we have $m_{x} =
m_{y} \approx\, 0.25 m_{z}$). These are clear and quantitatively
significant results emerging from our calculations. As far as we can
tell, they are ignored in all Landau-type dynamical investigations to
date.

Our results also reveal a considerable temperature dependence of the
damping constants or, equivalently, the time scale for
equilibration. Further, we obtain a marked anisotropy for the damping
in the ferroelectric phase, with $P_{z}$ fluctuations of the
spontaneous polarization equilibrating much faster than those in the
perpendicular plane. As far as we can tell, such effects are ignored
in all Landau-type dynamical simulations of ferroelectrics, despite
the fact that some of them (e.g., the general increase of damping
constants with temperature) are to be expected on basic physical
grounds. More exotic and unexpected effects, such as the found
anticrossing between the $z$-polarized soft phonon and an overtone in
the spectrum, are also unheard of in existing Landau-like or
effective-Hamiltonian treatments.

The present results thus suggest that it is worth examining the
assumptions implicit in dynamic simulations of soft-mode
ferroelectrics such as PTO, whether based on Landau-like theories or
coarse-grained effective Hamiltonians. It may be tempting to assume
that improvements over traditional approximations will not alter the
qualitative results, but this is not guaranteed. For example, our
preliminary studies of ferroelectric switching in BiFeO$_{3}$, based
on a time-dependent Landau approach, suggest that the resulting
switching path depends on the values of the damping constants
employed.\cite{robredomagro26b} Hence, there are good reasons to make
these approximations explicit and test them. Indeed, as Landau-like
simulations are increasingly regarded as quantitative and predictive
techniques, rather than merely phenomenological tools for physical
insight, such nuances will demand growing attention.

\subsection{Further methodological considerations}

We now comment on the generality and robustness of this approach to
calculate effective dynamic constants. More broadly, we discuss what
our investigation of PbTiO$_{3}$ suggests regarding our long-term
program for third-principles dynamic simulations of complex materials
at the mesoscopic and macroscopic scales.

\subsubsection{Comparison with a nonequilibrium approach}

The present approach has, at least, three important virtues. First, it
parallels the analysis of experimental spectroscopic data, which
allows us to benefit from decades of empirical know-how on the
topic. Second, it allows us to compute the full inertial and damping
tensors, at a given temperature, from a single equilibrium MD
simulation. Third, it naturally probes non-trivial anharmonic
phenomena (like overtones or overdamped modes) and allows us to
evaluate critically how realistic it is to treat the system in terms
of a single polarization vector $\Delta \bm{P}(t)$.

Having said this, one may wonder whether our reliance on equilibrium
dynamics, and the assumed harmonic nature of the relevant energy
landscape, may preclude an accurate quantification of effective
dynamic constants that, ultimately, we would like to apply in
nonequilibrium simulations. Indeed, it could seem that a more natural
strategy to tackle this problem would involve atomistic MD simulations
where we monitor the evolution of a perturbed state toward
equilibrium. This is, in essence, the approach adopted by Liu {\sl et
  al}. in Ref.~\onlinecite{liu21}. While such a procedure may present
difficulties of its own (e.g., getting access to the full $\bm{\mu}$
and $\bm{\gamma}$ tensors may require multiple simulations), it does
address explicitly the nonequilibrium processes of interest.

\begin{figure}
    \centering
   \includegraphics[width=\linewidth]{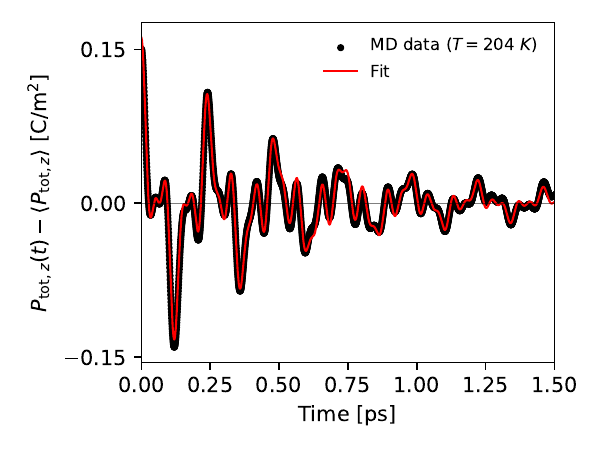}
   \caption{Time evolution of the polarization along the polar
     direction, $P_{{\rm tot},z}(t)-\langle P_{{\rm tot},z}\rangle$,
     during the relaxation from a perturbed initial state. The black
     line corresponds to the NVE trajectory computed at 204~K, while
     the red line shows the result of a fit to a model trajectory with
     three underdamped oscillators.}
    \label{fig:noneq}
\end{figure}

At the start of this project, we did consider such a nonequilibrium
approach to the calculation of effective dynamics
constants. Representative results are shown in
Fig.~\ref{fig:noneq}. We create a strongly perturbed initial state by
application of a large electric field along $z$ (10~MV/cm in the
example). We thermalize the system under such a field and for a given
temperature (200~K here). Then, at $t=0$, we remove the perturbation
and allow the system to evolve towards its zero-field equilibrium
state. (In this exercise we run a microcanonical MD simulation while
keeping the supercell lattice vectors constant, which results in a
small change of the initial temperature and other nuances, none of
them essential.) The computed $P_{{\rm tot},z}(t)$ can be fitted to a
superposition of three underdamped oscillators as given by
Eq.~(\ref{eq:Pt-underdamped}). Typically, the $P_{{\rm tot},z}(t)$
trajectory is dominated by a slow oscillator (our soft mode) for which
we can readily fit its characteristic frequency and damping
constant. In the example of Fig.~\ref{fig:noneq} we obtain $T=204$~K,
$\omega = 27.7$~THz, and $\gamma = 3.2$~THz for such a dominant
oscillator. For comparison, our equilibrium approach at $T=209$~K
yields $\omega = 25.9$~THz and $\gamma = 2.9$~THz, in remarkably good
agreement.

We observe this good correspondence between nonequilibrium and
equilibrium approaches in all our tests, suggesting that our method
yields effective constants that describe relaxation processes
accurately. Hence, on account of its simplicity and the advantages
mentioned above, we propose our equilibrium approach as the procedure
of choice for the calculation of effective dynamic constants.

\subsubsection{Alternative first-principles approaches}

Let us note that it is conceivable to devise alternative approaches to
obtain the effective dynamic constants of interest. Examples include
the family of self-consistent phonon methods used to compute
temperature-dependent phonon frequencies,\cite{hellman11,errea14} or
the schemes for lattice thermal conductivity studies based on the
Boltzmann transport equation;\cite{broido05} such schemes might be
adapted to obtain the frequencies and damping constants of specific
underdamped phonons, which would allow us to derive Landau-like
equations of motion for the corresponding order parameter. Also worth
noting are proposals to derive dynamical information from static
probability distributions (e.g., position-position and force-force
correlation functions)\cite{ljungberg13}, or even from diffraction
data.\cite{goodwin04,goodwin05} Yet, while these are valuable
approaches, the ongoing developments in the derivation of MLIPs
suggest that a direct calculation of the power spectrum will always be
accessible. Hence, we tend to favor the present method, which relies
on the exact MD trajectory of the system (save inaccuracies in the
MLIP or second-principles potential), thus avoiding approximations in
the treatment of phonon-phonon interactions.

\subsubsection{Several order parameters, lower symmetries}

Let us now imagine that we need to apply this procedure to a different
material. Here we discuss ferroelectric perovskite oxide BiFeO$_{3}$
(BFO) as a convenient example, but our considerations are general and
apply to arbitrary materials families and order parameters.

In the context of Landau-like approaches, we will always be interested
in describing the dynamics of one or more order parameters, or their
corresponding fields. In BFO's case, we have to address two primary
order parameters, the electric polarization $\bm{P}_{\rm tot}$
(essentially analogous to that of PTO) and the concerted antiphase
tilts of the oxygen octahedra that form the backbone of the perovskite
lattice. Such tilts can be described using an axial vector
$\bm{Q}_{\rm tot}$ whose components quantify the octahedral rotation
about each of the pseudocubic axes. The $\bm{P}_{i}$ contributions to
the total polarization can be considered to be decoupled, at the
harmonic level, from the $\bm{Q}_{i}$ contributions to the total tilt
vector (they transform according to different irreducible
representations of the reference space group $Pm\bar{3}m$). Hence, we
can write separate equations of motion for the dynamics of $\Delta
\bm{P}_{i}(t)$ and $\Delta \bm{Q}_{i}(t)$ oscillators around
equilibrium, of the form in Eq.~(\ref{eq:EoM-i}). (This statement is
strictly correct when the equilibrium state has cubic symmetry, but it
is not general. We come back to this point below.)

From an atomistic MD simulation of BFO at equilibrium, we can derive
both $\Delta \bm{P}_{\rm tot}(t)$ and $\Delta \bm{Q}_{\rm
  tot}(t)$. (The antiphase tilts are fully determined by symmetry in
the perovskite structure and, hence, computing their time evolution
from the MD trajectory boils down to a simple projection of atomic
displacements on a symmetry-adapted mode. A similar protocol would
apply to a general order parameter.) From these we can obtain the
corresponding power spectra, which we can in turn fit following the
procedure prescribed above. It is worth stressing here that the power
spectrum may not be available experimentally in the general
case. (This will typically be the case for BFO's tilts, in fact.)
However, computationally, calculating the spectrum for an arbitrary
order parameter is as easy as computing the dielectric one. Indeed, a
great advantage of a theoretical approach like ours is that we can
treat any order parameter of interest in essentially the same manner.

In our scheme, we need to be able to identify the spectral features
that we would like to capture in our Landau-like dynamical
simulations. As we have seen, even in relatively simple materials like
PTO, this assignment can get complicated by the presence of anharmonic
effects, including overtones and central peaks. Yet, identifying the
modes of interest becomes relatively simple in the limit of very low
temperatures, as we saw. The necessary simulations and analysis are
relatively trivial, irrespective of the complexity (e.g., the symmetry
and size of the unit cell) of the material under consideration. Hence,
we contend that this will be a straightforward step for the vast
majority of materials and order parameters. Note that, in difficult
situations, one might get information about the atomic displacements
associated with specific spectral features at finite temperature by
employing a more sophisticated analysis of the atomistic MD
trajectory.\cite{thomas10,rijal24}

Having said this, there are situations that may prove challenging for
a direct application of our methodology, and which we briefly address
now. First, our method relies on a decoupling, at the harmonic level
in the Landau potential $F$, of the order parameters of interest. In
the example of PTO, we have three independent polarization components
$\Delta P_{{\rm tot},x}$, $\Delta P_{{\rm tot},y}$, and $\Delta
P_{{\rm tot},z}$. It can be seen from symmetry that the harmonic
decoupling occurs in both the cubic ($\bm{P}_{\rm tot, eq} = \bm{0}$)
and tetragonal ($0 = P_{{\rm tot,eq},x} = P_{{\rm tot,eq},y} \neq
P_{{\rm tot,eq},z}$) phases. The simple form of the free-energy
landscape around such high-symmetry equilibrium states ensures a
trivial mapping between the power spectra $S_{{\rm
    tot},\alpha}(\omega)$ and the components $\Delta
P_{\alpha}(\omega)$ of the soft polarization, so that we can derive
their characteristic frequencies and damping constants from the
corresponding low-frequency underdamped phonon peaks. In the general
case, though, the situation may be more intricate.

For example, in BFO's case the mentioned harmonic decoupling applies
to the cubic phase of the material (with $\bm{P}_{\rm tot,eq} =
\bm{Q}_{\rm tot,eq} = \bm{0}$), where $F$ does not present any crossed
harmonic interactions of the form $\Delta P \times \Delta Q$. However,
in BFO's ferroelectric phase, where both polarization and tilts have
non-zero equilibrium values, symmetry allows for $\sim \Delta P \times
\Delta Q$ couplings. In such a situation, some of the soft phonons of
interest will in fact have a mixed polar and tilt character. In turn,
the frequencies and damping constants that can be derived from the
corresponding peaks in the power spectra will not correspond exactly
to a soft polarization or a soft tilt, but to a certain linear
combination of the original order parameters. An elegant treatment of
this situation would require knowledge of the eigenvectors of such
mixed soft phonons -- which could be approximated from phonon
calculations at 0~K, or derived from the atomistic MD trajectory
through a more involved analysis --, so that one can recover a
harmonic decoupling through a suitable change of basis. Nevertheless,
we suspect that, in practice and in most cases, the mixing of order
parameters will be relatively weak, so that we will retain phonons of
a dominant character (mainly $P$-like or mainly $Q$-like in BFO's
example) and a direct application of the method presented here will
yield sufficiently accurate dynamic constants for the different order
parameters. This question remains for future investigation.

\subsubsection{Static vs. dynamic Landau approaches}

In a recent publication, we have introduced a machine-learning method
to derive Landau-like free-energy potentials for macroscopic (and,
eventually, mesoscopic) simulations.\cite{pulzone25} Such a
``third-principles'' scheme is well-defined and seemingly
straightforward to apply to any material and combination of order
parameters, and our group is already working on potentials for other
compounds.

The present work tackles the extension of the third-principles
approach to the time domain. Our results for PTO are satisfactory, in
the sense that they provide us with physically meaningful dynamic
constants that will allow us to run predictive simulations of this
compound's nonequilibrium polarization dynamics, using a Landau-like
approach that reproduces accurately atomistic results much more costly
to obtain computationally. Nevertheless, the time-dependent problem
presents difficulties that are not present in its static
counterpart. Let us discuss them briefly, using the polarization
$\bm{P}$ as a convenient example of order parameter, noting that our
considerations apply generally.

The main difficulty pertains to the difference between $\Delta
\bm{P}_{\rm tot}$, which includes all distortions with a polar
symmetry, and $\Delta \bm{P}$, which is the soft part of the
polarization that dominates the structural phase transitions and
related properties. Let us stress that, when it comes to computing the
static equilibrium properties of a system, including the response to
static external fields, the internal structure of $\Delta \bm{P}_{\rm
  tot}$ (as given by Eq.~(\ref{eq:multimode})) can typically be
ignored. However, if we want to investigate the response to
time-dependent perturbations, as e.g. field pulses that essentially
span the whole frequency spectrum, we need to worry about the $\Delta
\bm{P}_{i}(t)$ components adding up to $\Delta \bm{P}_{\rm
  tot}(t)$. Note that any particular contribution $\Delta
\bm{P}_{i}(t)$ can become resonant and dominate the physics if the
system is probed at a frequency close to $\omega_{i}$. In such a case,
a dynamic description based on a simple Landau-like model, considering
a single low-frequency oscillator $\bm{P}$, or a single monolithic
variable $\bm{P}_{\rm tot}(t)$, would be obviously inappropriate. Let
us stress that these considerations are relevant in many cases of
current interest, including nonequilibrium {\it phononic} approaches
where high-frequency modes are strongly excited to modify, via
anharmonic couplings, the free-energy landscape relevant to low-energy
phenomena (e.g., the ferroelectric switching of spontaneous
polarization).\cite{forst11,subedi14,chen22} In passing, note that
this issue also affects typical coarse-grained effective Hamiltonians
where the total polarization is approximated by its soft part.

Whenever we need to model several dynamically distinct contributions
to the total polarization, we must abandon a simple Landau potential
$F(\bm{P}_{\rm tot})$, even if such a model may be sufficient for
static investigations. Instead, we need to use a more complex
$F(\bm{P}_{1},\bm{P}_{2},...)$ with suitably chosen polar
variables. For example, here $\bm{P}_{1} = \bm{P}$ could be the soft
part of the polarization, while $\bm{P}_{2}$ could correspond to a
high-frequency polar mode that we plan to excite. Note that this
extension does not introduce any fundamental difficulty. Existing
methods to compute $F(\bm{P}_{\rm tot})$ -- as the one some of us
introduced in Ref.~\onlinecite{pulzone25} -- can be easily adapted to
map the free-energy landscape as a function of $\bm{P}_{1}$ and
$\bm{P}_{2}$, including all relevant cross couplings, by simply
treating them as separate order parameters. Knowledge of the atomic
motions underlying such order parameters would be required, but we
have access to this information -- usually to a very good
approximation -- through a phonon calculation at 0~K. Further, our
present method to derive dynamic constants can be readily applied as
long as we are able to identify the spectral peaks corresponding to
$\Delta \bm{P}_{1}$ and $\Delta \bm{P}_{2}$. This problem is not
fundamentally different from that of BFO, discussed above. In BFO's
case, we need to treat two order parameters of different symmetries,
while now we are considering the possibility of modeling two order
parameters with the same symmetry; this does not introduce any
difficulty.

\subsubsection{Practical considerations on using our results}

Now we briefly comment on how to use the information presented in this
article in order to run Landau-like dynamic simulations of
PbTiO$_{3}$, our considerations being applicable to the general case.

As we have seen, we find that at most temperatures PTO can be
described using a single polarization vector $\bm{P}$ whose components
behave as underdamped oscillators. We have computed characteristic
inertial and viscous-damping constants for each polarization component
and temperature. The obtained values can be directly used in a
Landau-like dynamical simulation.

We find, though, that most of these constants do not present a simple
temperature dependence. On the one hand, we observe a discontinuity at
$T_{\rm C}$, reflecting a non-trivial dependence on the equilibrium
state of the system. This includes a sizeable anisotropy in the
inertial and damping tensors, an effect that will probably be present
in all compounds undergoing a structural phase transition, with a
magnitude proportional to that of the symmetry-breaking distortion. On
the other hand, as shown in Figs.~\ref{fig:fits} and \ref{fig:mass},
the anticrossing observed for $\Delta P_{z}(\omega)$ around 400~K
results in a nonmonotonic behavior of $\gamma_{z}$ (and $c'_{z}$) and
a discontinuity in $m_{z}$ (and $\mu_{z}$). We thus have evidence
that, if we want to reproduce the atomistic results closely, it will
be necessary to account for the intricate temperature dependence of
the dynamic constants, and even extend the Landau-like potential to
include a polar overtone below $T_{\rm C}$. Alternatively, if we just
want to account for the main effects -- i.e., the anisotropy of the
dynamical behavior and its basic temperature dependence --,
implementing the fitted functions given in Table~\ref{tab:fits} could
be enough. In the absence of specific reasons to complicate the
treatment, we suggest this pragmatic approach.

Finally, there are temperature ranges where (some of) our soft polar
oscillator becomes overdamped. To simulate the system in such
conditions, we need to make $\mu_{\alpha} = 0$ in the corresponding
equations of motion and use the viscous-damping constants fitted to
the corresponding central peaks (see Fig.~\ref{fig:fits}(b)).

\subsubsection{Time- or state-dependent dynamic constants}

Let us now comment on a second practical consideration that may be
less obvious at first, but which is critical even for a material as
simple as PbTiO$_{3}$. Our results clearly show that inertial and
damping constants are anisotropic, the polarization oscillations along
the polar axis ($\Delta P_{z}$ in our presentation) being dynamically
distinct from those occurring in the perpendicular plane ($\Delta
P_{x}$ and $\Delta P_{y}$). Imagine now that we want to investigate
the action of an electric field $\mathcal{E}_{x}>0$, driving the
system from a $z$-polarized state to an $x$-polarized one. To run such
a simulation, we should obviously account for the fact that the
dynamic constants must adapt to the polarization rotation. Considering
the viscous damping $\bm{\gamma}$ as an example, we might work with
the approximate expression
\begin{equation}
 \gamma_{\alpha}(T,\bm{P}(t)) = \gamma_{\alpha}(T) +
 \sum_{\beta\delta} \gamma_{\alpha,\beta\delta}(T) P_{\beta}(t)P_{\delta}(t)
 + ... \; ,
 \label{eq:gamma-P}
\end{equation}
where we use the fact that the cubic symmetry of the perovskite
reference structure precludes a linear dependence on the
polarization. Further, the only non-zero harmonic terms are
$\gamma_{x,xx} = \gamma_{y,yy} = \gamma_{z,zz}$ and $\gamma_{x,yy} =
\gamma_{x,zz} = \gamma_{y,xx} = \gamma_{y,zz} = \gamma_{z,xx} =
\gamma_{z,yy}$. Analogous expressions can be written for all the other
dynamic tensors introduced in this work: $\bm{\mu}$, $\bf{c}$, and
$\bf{c}'$.

It is worth noting that such a detailed modeling of the dynamic
constants would provide us with a quantitative understanding of how
they depend on the actual magnitude of the polar distortions, thus
going beyond the mere directional dependence evidenced in
Figs.~\ref{fig:fits}(b) and \ref{fig:mass}. Such considerations could
be relevant in many contexts, as e.g. in time-dependent
Ginzburg-Landau simulations to study the response of ferroelectric
domain walls, where the polarization state sharply differs from that
of the domains. (For such phase-field studies, we would make the
dynamic tensors dependent on the local instantaneous polarization
$\bm{P}(\bm{r},t)$.) Indeed, these issues will be a focus of attention
for us in the future, in particular when extending our
third-principles program to the systematic fitting of dynamic
Ginzburg-Landau models.

We should note that these considerations are not entirely new. For
example, Durdiev {\it et al.} have recently used atomistic MD
simulations of domain-wall motion in BaTiO$_{3}$ to derive a
time-dependent Ginzburg-Landau theory, using damping constants
(related to the ``mobility coefficients'' in their article) that
account for the crystallographic anisotropy.\cite{durdiev25}

\subsubsection{Toward third-principles dynamical simulations}

Taking the cue from the last comment, let us conclude with a broader
reflection on developing predictive third-principles methods for
nonequilibrium dynamical simulations at the mesoscopic or macroscopic
scales of arbitrary materials.

The approach described here provides clear guidelines for how to
proceed and no fundamental obstacles appear insurmountable. Yet, it
remains an open question whether additional simplifications will be
required to design a procedure for deriving a reduced set of dynamical
variables, and their associated inertial and damping constants, in the
general case. In particular, we believe that, at present, it would be
premature to attempt a fully automated, machine-learning-style
implementation of this workflow. Even if we restrict ourselves to the
perovskite family, a broader set of representative case studies will
be necessary to assess whether the present methodology (and its
anticipated extensions outlined above) is sufficient, or whether
further conceptual developments or practical approximations will be
required.

It is worth emphasizing again the specific challenges of the
time-dependent problem. Static equilibrium behavior can often be
described by a single order parameter, $\bm{P}_{\rm tot}$ for the
ferroelectric compound considered here. However, a dynamical
description may require resolving contributions to $\bm{P}_{\rm tot}$
associated with different characteristic frequencies. Consequently,
Landau or Ginzburg–Landau models that suffice for investigating
quasi-static phenomena may be ill-suited to provide a dynamical
picture that is quantitatively accurate. Rather, they will typically
need to be extended to include all relevant, dynamically distinct
degrees of freedom. To our knowledge, this requirement to work with
more complete Landau-like models has not been generally recognized in
the literature. Yet, the complexity already observed in simple lead
titanate strongly suggests that such extensions will be essential for
developing predictive third-principles theories, particularly if they
are to be derived through automated machine-learning approaches.

We may wonder whether the case of PbTiO$_{3}$ considered here may make
our third-principles program seem more trivial (or more involved) than
it really is. In this regard, one should note that the simplicity of a
material is essentially determined by the number of atoms in the unit
cell, which largely controls how many dynamically distinct variables
will contribute to the time evolution of a given order parameter. PTO
has a modest five-atom equilibrium unit cell at all temperatures and a
relatively small amount of polar phonons. Then, in light of the
multiplicity of non-trivial dynamical effects observed in this case,
it is clear that compounds and phases with bigger cells can be
expected to present rich and intricate spectra, which should
complicate the task of deriving accurate dynamic theories.

Having said this, we should also note that, despite their small unit
cells, perovskite oxides like PTO are well-known for hosting many
low-energy phonon bands that yield intricate structural and
lattice-dynamical behaviors, including highly anharmonic effects. In
other words, these materials are not as simple as they may seem. By
contrast, our past experience with some novel ferroelectrics -- such
as hexagonal boron nitride, a classic van der Waals
compound\cite{wang24} -- suggests that their dynamical properties
could be amenable to a relatively simple Landau-like description, and
that a straightforward application of the current approach could prove
sufficient in that context.

\section{Summary}

We have presented the foundations of a method to derive models for
predictive simulations of nonequilibrium dynamical phenomena at
mesoscopic and macroscopic scales. We show that such a
``third-principles'' dynamical approach can be derived from
first-principles atomistic simulations, enabling the investigation of
arbitrary materials and order parameters in a way that is largely
independent of experimental constraints. For example, our scheme
allows the study of dynamical features that are difficult to probe
experimentally (e.g., antipolar phonons) as no conjugate fields are
readily available.

We illustrate the method with an application to perovskite oxide
PbTiO$_{3}$, a prototypical soft-mode ferroelectric that provides a
formally simple yet physically challenging test case. We also discuss
how the results can be used in practical dynamical Landau or
Ginzburg–Landau simulations, including a critical assessment of open
issues requiring further work.

It is worth stressing that we rely on an analysis of equilibrium
dynamics, extracting the full inertial and damping tensors from the
spontaneous excitations and equilibration processes of the modes of
interest. Our numerical tests suggest that this is sufficient to
accurately characterize the time scale over which the system relaxes
to equilibrium after relatively strong perturbations.

In conclusion, we hope this work will advance predictive mesoscale and
macroscale simulations of complex dynamical and nonequilibrium
phenomena, of both fundamental interest and technological relevance
for applications exploiting ultrafast, low-energy control of nonlinear
materials in next-generation electronics and computing. This remains a
vast undertaking. We expect the present study to contribute by
providing a general framework that will enable further developments,
including the eventual automated generation of models through
machine-learning approaches.

\section*{Acknowledgments}

We thank L.~Bellaiche (Arkansas), A.~Gr{\"u}nebohm (Bochum), and
J.~Hlinka (Prague) for many fruitful discussions over the years. Work
supported by the Luxembourg National Research Fund (FNR) through
project C21/MS/15799044/FERRODYNAMICS.

\section*{Data availability}

The data that support the findings of this article are openly
available at Ref.~\onlinecite{data}.


%

\end{document}